\documentclass[journal]{vgtc}                     

\onlineid{0}
\vgtccategory{Research}

\title{A Multi-Level, Multi-Scale Visual Analytics Approach to Assessment of Multifidelity HPC Systems}

\author{Shilpika, Bethany Lusch, Murali Emani, Filippo Simini, Venkatram Vishwanath, Michael E. Papka, and Kwan-Liu Ma}

\authorfooter{
  \item
  	Shilpika is with University of California, Davis \& Argonne Leadership Computing Facility.
  	E-mail: fshilpika@ucdavis.edu
  \item
  	Bethany Lusch is with Argonne Leadership Computing Facility, 
Argonne National Laboratory.
  	E-mail: blusch@anl.gov.
   \item
  	Murali Emani is with Argonne Leadership Computing Facility, 
Argonne National Laboratory.
  	E-mail: memani@anl.gov.
   \item
  	Filippo Simini is with Argonne Leadership Computing Facility, 
Argonne National Laboratory.
  	E-mail: fsimini@anl.gov.
   \item
  	Venkatram Vishwanath is with Argonne Leadership Computing Facility, 
Argonne National Laboratory.
  	E-mail: venkat@anl.gov.
   \item
  	Michael E. Papka is with Argonne Leadership Computing Facility at 
Argonne National Laboratory \& 
Department of Computer Science at
University of Illinois Chicago
  	E-mail: papka@anl.gov.
    \item
  	Kwan-Liu Ma is with Department of Computer Science at
University of California, Davis\\
  	E-mail: klma@ucdavis.edu.
}

\abstract{
The ability to monitor and interpret of hardware system events and behaviors are crucial to improving the robustness and reliability of these systems, especially in a supercomputing facility. The growing complexity and scale of these systems demand an increase in monitoring data collected at multiple fidelity levels and varying temporal resolutions. In this work, we aim to build a holistic analytical system that helps make sense of such massive data, mainly the hardware logs, job logs, and environment logs collected from disparate subsystems and components of a supercomputer system. This end-to-end log analysis system, coupled with visual analytics support, allows users to glean and promptly extract supercomputer usage and error patterns at varying temporal and spatial resolutions. We use multiresolution dynamic mode decomposition (mrDMD), a technique that depicts high-dimensional data as correlated spatial-temporal variations patterns or modes, to extract variation patterns isolated at specified frequencies. Our improvements to the mrDMD algorithm help promptly reveal useful information in the massive environment log dataset, which is then associated with the processed hardware and job log datasets using our visual analytics system. Furthermore, our system can identify the usage and error patterns filtered at user, project, and subcomponent levels. We exemplify the effectiveness of our approach with two use scenarios with the Cray XC40 supercomputer.
}


\teaser{
  \centering
  \includegraphics[width=1\textwidth]{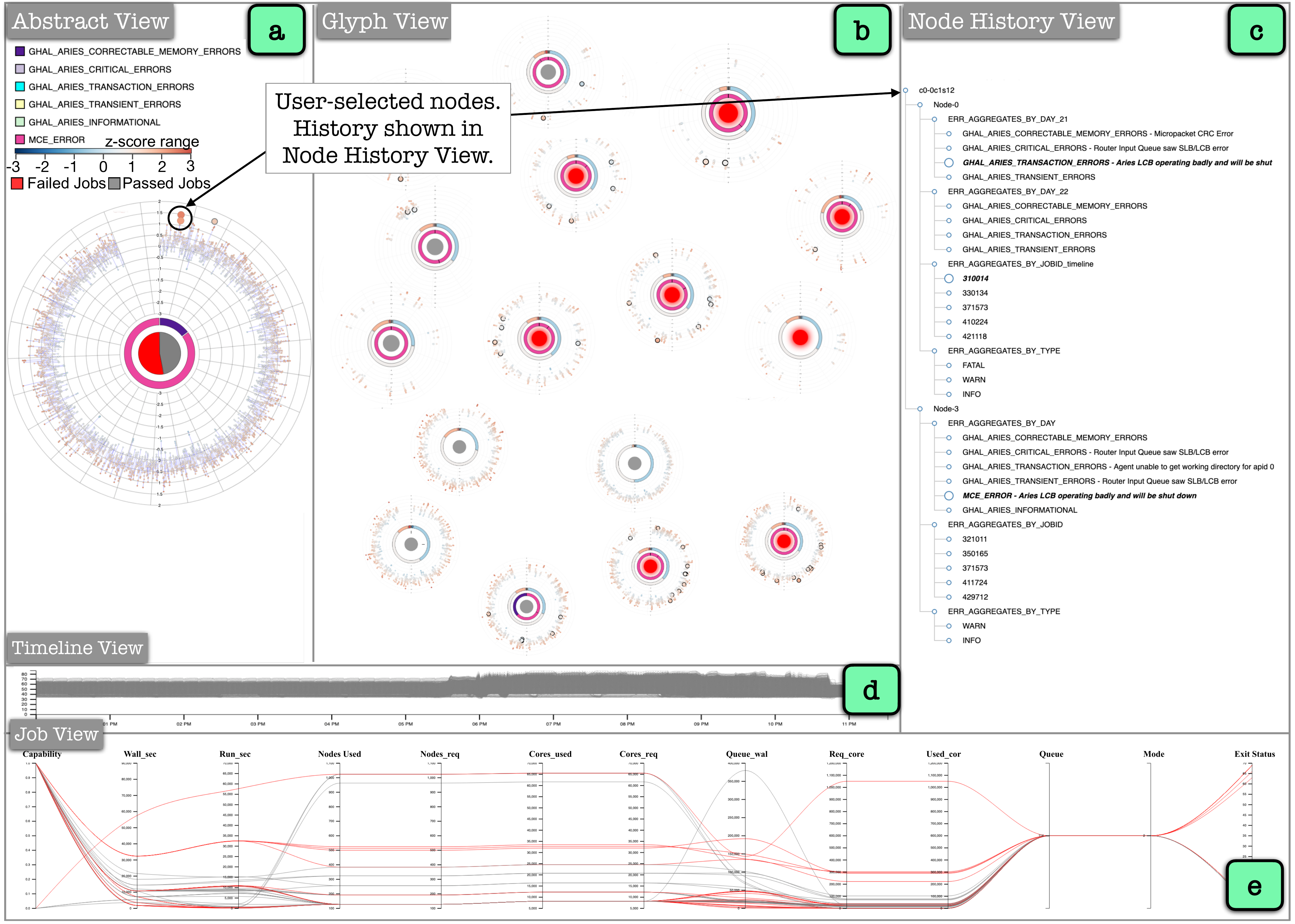}
  \caption{
  	The UI of our visual analytics tool. (a) The abstract view shows the aggregated values of the log data. (b) The glyph view shows the log data information separated by granularities of jobs, users, projects, nodes used, and job exit codes. (c) The node history view shows the history of the nodes selected by lasso selection in the abstract and the glyph views. (d) The environment log data (power, temperature, voltage, current, etc.) is displayed in the timeline view. (e) The job view uses parallel coordinates to display the details of the job log information.
  }
  \label{ui}
}

\graphicspath{{figs/}{figures/}{pictures/}{images/}{./}} 
\usepackage{dblfloatfix}
\usepackage{tabu}                      
\usepackage{booktabs}  
\usepackage{amsmath, xparse}
\usepackage{lipsum}                    
\usepackage{mwe}                       
\usepackage{mathptmx} 
\newcommand{\mathrmbf}[1]{\mathrm{\mathbf{#1}}}
\usepackage{bm}
\usepackage{amsfonts}
\usepackage{amssymb}
\usepackage{stackengine}

\begin{document}

\maketitle
\firstsection{Introduction}
\label{sec:Intro}

Maintenance and monitoring of supercomputer systems are crucial steps for facilitating their robustness and reliability, leading to advancements in numerous fields of scientific research. As supercomputing systems are growing in complexity with changing execution environments, increasing computation capabilities, and increasing frequency of updates and upgrades, the monitoring data collected at multiple fidelity levels and
varying temporal resolutions is also increasing. Furthermore, high utilization of these systems is expected to accommodate applications and jobs that sometimes execute on these systems for weeks. Hence, any system errors and failures could compromise the integrity of the results or cause job failures, all of which lead to significant overhead in the already computationally and financially expensive research and development. 
Therefore, we study these diverse systems and subsystems’ log data and use the knowledge acquired to derive patterns of behaviors that could help understand the underlying state of the system at any given time. In this paper, we focus our efforts on a Cray XC40 supercomputer system. The nodes are interconnected with an Aries interconnect in a Dragonfly topology. We built a holistic analytical system that processes the massive log data, specifically the (i) hardware logs (event sequence and text data),  (ii) job logs (event sequence data), and (iii) environment logs, also referred to as SEDC (System Environment Data Collections) logs. SEDC is a Cray systems tool used to collect and report temporal data (readings collected from sensors housed in the compute nodes) in real-time, collected from disparate subsystems and components of a supercomputer system. This data collected from the predefined spatial localities and various temporal resolutions are unprocessed and raw. Therefore, we deal with multifidelity large-scale data from various sources within the HPC system. 
For example, the hardware error log contains information accumulated from different control systems and interlinked subsystems with data size ranging in tens of GB. The environment log data are more frequently reported and collected at every 10-30 seconds interval, making the dataset size approach gigabyte (GB) to terabyte (TB) range every few weeks. The job log data contains information about the various applications utilizing the systems and their characteristics (i.e., nodes used, start and end times, etc.) with data sizes ranging in hundreds of megabytes (MB) a year. The kinds of logs range in size from 5 GB per day. It is a challenging task to analyze these volumes of data daily.

 This work uses multiresolution Dynamic Mode Decomposition (mrDMD)~\cite{mrDMD} to succinctly represent the larger environment log data. We then display its correspondence with the additional preprocessed log data like the hardware and job log data. MrDMD is a technique that decomposes high-dimensional data into correlated spatial-temporal modes and is an improved version of Dynamic Mode Decomposition (DMD) ~\cite{dmd_2009, schmid_2010, Schmid2011}. DMD decomposes the data into spatial modes that correlate it over its spatial features (similar to principal component analysis (PCA)) and temporal modes that correlate the data across unique temporal Fourier modes. The DMD algorithm has been used to isolate and extract distinct sleep spindle networks from sub-dural electrode array recordings of human subjects using DMD spectrum analysis and baselines analysis~\cite{BRUNTON20161}. Furthermore, mrDMD extracts the spatial and temporal features over multiple timescales by integrating space and time. The multiresolution analysis in mrDMD recursively subtracts low-frequency, or slower varying, dynamics from the data at each selected timescale, making it ideal for separating different timescale features or modes and analyzing them separately. MrDMD has been used for object feature tracking in videos over evolving timescales~\cite{mrDMD}. 
 
Our work combines the mrDMD analysis~\cite{mrDMD} with the work on frequency isolation using DMD spectrum analysis to generate the mrDMD spectrum~\cite{BRUNTON20161}. We extract the high power mrDMD modes at various frequency ranges. We then identify custom baselines that are either system or user-specific to extract patterns that signify changes between the typical system state (specified by the baselines) and the current state. 

We make the following contributions:
\begin{itemize}
\item MrDMD Baseline identification for system logs is based on two factors
\begin{itemize}
    \item system or sub-component specification
    \item user system-usage trends
\end{itemize}
\item Our mrDMD algorithm has the following improvements and features

\begin{itemize}
    \item Time range splits at mrDMD levels are determined by the start and end times of system jobs. This ensures that the results of the system usage analysis do not overlap between different jobs or applications utilizing the system.
    \item MrDMD mean mode magnitudes are extracted at frequency ranges where the ``power'' of the corresponding mode is higher than a specified threshold, thus eliminating noise.
    \item We process the time range splits asynchronously at each level. This improvement increases the speed of the algorithm. 
\end{itemize}

\item Two use scenarios on real-world datasets demonstrating the effectiveness of our approach.
\end{itemize}

Since the data is proprietary, compiling this diverse multifidelity dataset from other supercomputing facilities is challenging. Nevertheless, our stated contributions will apply to other large-scale systems reporting similar data types. In addition, it will provide insights to users and system administrators to better understand the system's state at any given time, present or past, and to perform system maintenance to reduce future system failures proactively. 

\section{Related Work}
\label{sec:RelatedWork}
\subsection{Error Log Analysis in Large Scale Systems}
Substantial efforts have been directed towards pinpointing and forecasting failures to facilitate evasive actions for failure identification in large-scale HPC systems, especially when current technology and infrastructure are delivering the capacity to record systems' states faster. Past survey papers~\cite{b32,b33,surveyreliability} on forecasting and classifying failures give an in-depth view of current methodologies and their drawbacks based on log data analysis. 

Pre-processing of log data is the first step in log analysis, and the steps are grouped into three main actions: data categorization, data filtering, and causality filtering ~\cite{b13}. Some past efforts include real-time clustering algorithms to concisely represent temporal and event-based data with evolving clusters of information in the log files~\cite{b31, 9751445, b8,b14, 9825952}. 

Past efforts also include visual analytics solutions for studying large-scale system behavior. Some previous works include developing a multi-coordinated visual analytics tool on Dragonfly network topology to investigate and optimize the network communication or system behavior of a supercomputer ~\cite{FUJIWARA201898, 8048931, shilpika2019mela, 9825952}, developing node-link diagrams with matrix-based hierarchical aggregation techniques to visualize any network topology~\cite{8585646}, functional data analysis (FDA) to incrementally and progressively update the streaming time series data collected from hardware systems with a focus on identifying outliers by using FDA~\cite{9751445}, and a visual analytics framework with automated time-series analysis and visualizations for analyzing HPC datasets produced by parallel discrete-event simulations (PDES)~\cite{8973380}. Text-based system logs can be system-generated or user-generated by an administrator or operator. These logs may contain varying abbreviations and admin-specific terminologies. Visual analytic approaches have been proposed to help filter and analyze millions of records of such inconsistent text data~\cite{10.1145/3411764.3445396,255601}. Word occurrence frequency and information-theoretic metrics are used to visually highlight common and uncommon issues and fixes that occur in the maintenance logs~\cite{255601}. Past works on text-based analysis display content and structure using features such as term frequencies, co-occurrences, and sentence structures. ConceptScope,~\cite{10.1145/3411764.3445396}, provides conceptual overviews incorporating domain knowledge using Bubble Treemap visualization, multiple coordinated views of document structure, and a hierarchy of concepts with data overviews. A high-level abstracted overview of CCTs (Calling Context Tree) using interactive visual analytics tool~\cite{8901998} have also helped handle large event-based log data. CCTs help understand the execution and performance of parallel programs using performance metrics with call paths. With semantic refinement operations, performance bottlenecks are progressively explored on applications running on multiple parallel processors. EnsembleCallFlow~\cite{9622132} supports the exploration of ensembles of call graphs, a combination of performance metrics and application execution contexts. They introduce ensemble-Sankey, combining resource-flow (a Sankey plot to describe the graphical nature of the call graph) and box-plot (to convey the performance variability within the ensemble) visualization techniques. A few automated log analysis tools~\cite{b2,b8,b13,b14,b15,b16} use text or event pattern-based correlations, signal processing, pattern recognition and mining, and spatiotemporal event-based analysis. LogAider~\cite{b2} and LogMaster~\cite{b18} include generic, easy-to-use visualizations that can extract event correlations. These tools analyze a specific type of data or a combination of data types, e.g., numerical, text-based, or event-based data. Our log data contains all three data types; thus, the task becomes challenging to process these data types individually and in conjunction. The results of the analysis are displayed on our visual analytics tool. 

DCDB Wintermute~\cite{b92} is a generalizable framework using multiple types of log data, including job, hardware, and environment logs, and implemented on the DCDB (Data Center Data Base) monitoring system. While this work requires integrating the online ODA (Operational Data Analysis) framework in an HPC system, our work relies on data pre-processing of raw data considering the types of logs and their correspondence with each other. Past efforts also include scalable and interactive visual analytics tools~\cite{shilpika2019mela, 9825952, kesavan2020visual} for analyzing multifarious log data; however, these tools lack the capability to analyze and visualize extensive temporal data in a concise representation. Legacy tools like IBM Blue Gene Navigator~\cite{b11} with plain log statistical visualization are popular tools operational administrators use to monitor supercomputer health.  

In this work, we have a front-end visualization showing correlations between hardware log, job log, and environment log data with time visually synchronized. In order to provide a holistic insight into comprehending the complexity of large-scale systems, we need to process and analyze a multitude of log data. Past attempts typically process only one or two types of error log data, and most studies ignore the analysis of the larger and more computationally expensive environment logs. In our work, we aim to address these shortcomings.

\subsection{Multiresolution Dynamic Mode Decomposition}
As mentioned above, mrDMD is a variation of the DMD algorithm. The original DMD algorithm was applied to fluid dynamics. It works with the assumption that observations can be approximated by a linear dynamical system that closely models the characteristics of the observations~\cite{mrDMD}. Therefore, DMD and mrDMD extract information from a nonlinear dynamical system~\cite{dmd_2009, schmid_2010, mrDMD}. Inspired by multiresolution analyses and, specifically, wavelet methods and windowed Fourier transforms, mrDMD recursively analyzes timescales, selecting modes to remove from the data of interest~\cite{mrDMD}. 

In recent years both DMD and mrDMD have served as powerful mechanisms for studying the dynamics of the nonlinear systems in fields including fluid mechanics~\cite{schmid_2010, mrDMD, tut}, financial analysis~\cite{Mann2015DynamicMD}, control systems~\cite{cntrl} neuroscience~\cite{BRUNTON20161}, streaming analysis~\cite{Hemati2014DynamicMD, Pendergrass2016StreamingGS} and denoising~\cite{Dawson2014CharacterizingAC, Schmid2011}, and foreground and background separation in video analysis~\cite{mrDMD, Manohar2019OPTIMIZEDSF}. DMD with unsupervised clustering was used to uncover distinct sleep spindle networks using cortical distribution patterns, frequency, and duration~\cite{BRUNTON20161}. DMD has also been extended to incorporate control and demonstrated on a model with applicability to infectious disease data analysis with mass vaccination~\cite{cntrl}. MrDMD of elastic waves with image registration and Kullback Leibler (KL) divergence was used to diagnose and localize the surface microscale defects in the Lead Zirconate Titanate~\cite{9524582}. MrDMD has been used to study and detect the onset of seizures in scalp electroencephalography (EEG) signals~\cite{8709782}. In a streaming low-storage fluid dynamic setting, DMD results were updated as new data became available. Their methods included a "batch-processed" formulation and a compression step~\cite{Hemati2014DynamicMD}. Another parallelized algorithm extracts DMD modes using the streaming method of snapshots singular value decomposition on a graphics processing unit (GPU). They use the native compressed format of many data streams to reduce data transfer costs from CPU to GPU~\cite{Pendergrass2016StreamingGS}. A denoising DMD implementation to reduce the algorithm's bias to sensor noise used three steps, including correcting the determined bias using known noise properties, performing DMD forwards and backward in time, combining the results, and developing an algorithm based on least-square analysis~\cite{Dawson2014CharacterizingAC}.

Since DMD and its variations have numerous applications in various fields, we attempt to leverage its benefits by combining the mrDMD analysis~\cite{mrDMD} with the DMD spectrum analysis for isolating mrDMD modes at automated or user-specified frequency ranges~\cite{BRUNTON20161}. We have improved the sampling rate selection and temporal split identification in the recursion steps of the mrDMD algorithm. At each level of the multiresolution analysis, we compute mrDMD modes asynchronously which further increases the speed of our algorithm. These improvements increase the accuracy and speed of computation. We also choose baselines that we define as system and user-specific to extract meaningful patterns that indicate changes between the typical system state (represented by baselines) and the current state (represented by data of interest). To the best of our knowledge, currently, no tools visualize the mrDMD analysis results. Our tool visualizes the z-scores or standard deviation of the mrDMD results of current readings from the mrDMD results of chosen baseline readings.
\section{System Overview}
\label{sec:SystemOverview}
This section describes the overview of our visual analytics tool and the back-end analysis pipeline.

\subsection{Overall Organization}

\begin{figure}[ht!]
\centerline{\includegraphics[width=\linewidth]{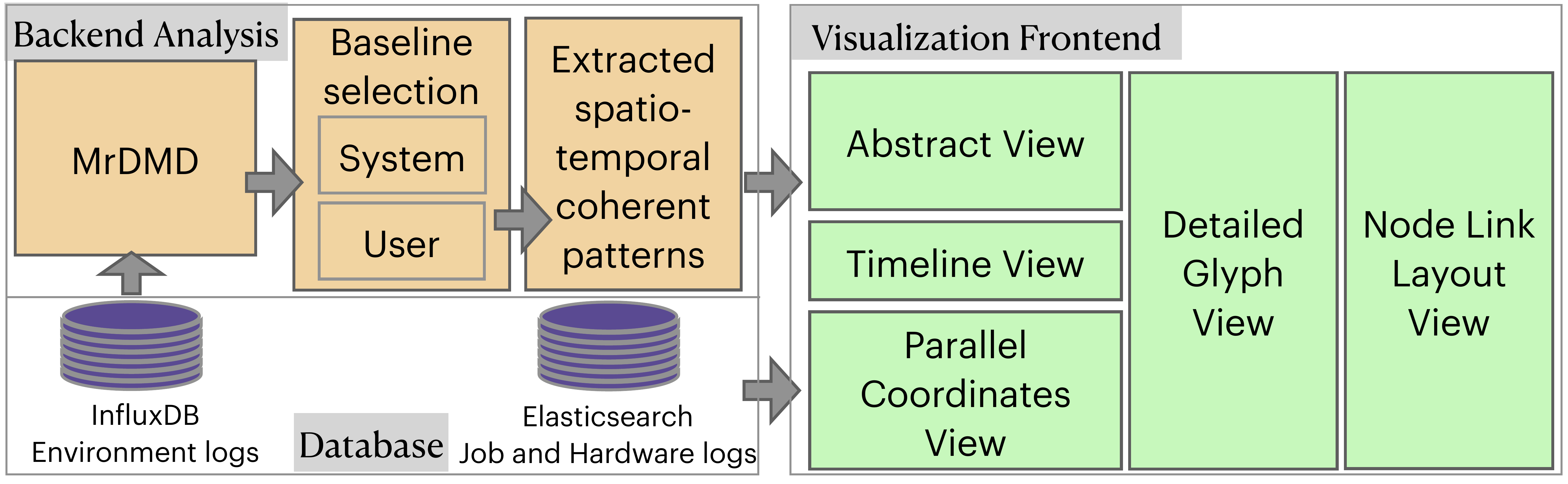}} 
\caption{General organization of our visual analytics tool}
\label{ao}
\end{figure}

Figure~\ref{ao} depicts the general organization of our system with three main sections: databases, back-end analysis, and front-end visualization. Our system operates on two 6-core, 2.4-GHz Intel E5–2620v3 processors (Intel Haswell architecture) and 256GB of DDR4 memory. We use InfluxDB~\cite{b6}, a time series database optimized for speedy storage and retrieval of time series data to store environment logs. The job and hardware logs are stored in an Elasticsearch database~\cite{esearch}. The data sizes are 3-4 GB per day, 1-2 GB per month, and a few hundred MB per year for the environment, hardware, and job logs. 
Our system operates on Python 3.8, a Flask server~\cite{b46}, and D3~\cite{b47} visualization. The source code will be made available on publication. The back-end analysis section is split into three groups consisting of the mrDMD analysis with our improvements, the baseline selection mechanism  (where we choose between user-specific or system-specific baselines), and the extraction of the coherent spatiotemporal patterns based on the mrDMD power spectrum values. The visual analytics frontend consists of five sections with a built-in interaction mechanism (explained in detail in Fig.~\ref{ui}).

\subsection{Backend Analysis}

Jobs routinely utilize supercomputers for hours to weeks on thousands of nodes. A supercomputer node reports hardware and software errors along with information about jobs utilizing it. Each node also has multiple sensors that record power, temperature, current, memory, fan speed, etc., and are stored as environment logs. Our backend analysis pipeline mainly focuses on processing this large environment log dataset collected from sensors. This log data contains monitoring information from different sensors collected every $10$-$30$ seconds. The resulting substantial data size poses a challenge and contributes to significant processing overhead. The supercomputer houses approximately 150 sensor readings per node. Furthermore, this dataset is usually ignored in log data analysis due to the size and lack of valuable insights extracted from popular methods. 

\subsubsection{Multiresolution Dynamic Mode Decomposition(mrDMD)}

\begin{figure}[ht!]
\centerline{\includegraphics[width=\linewidth]{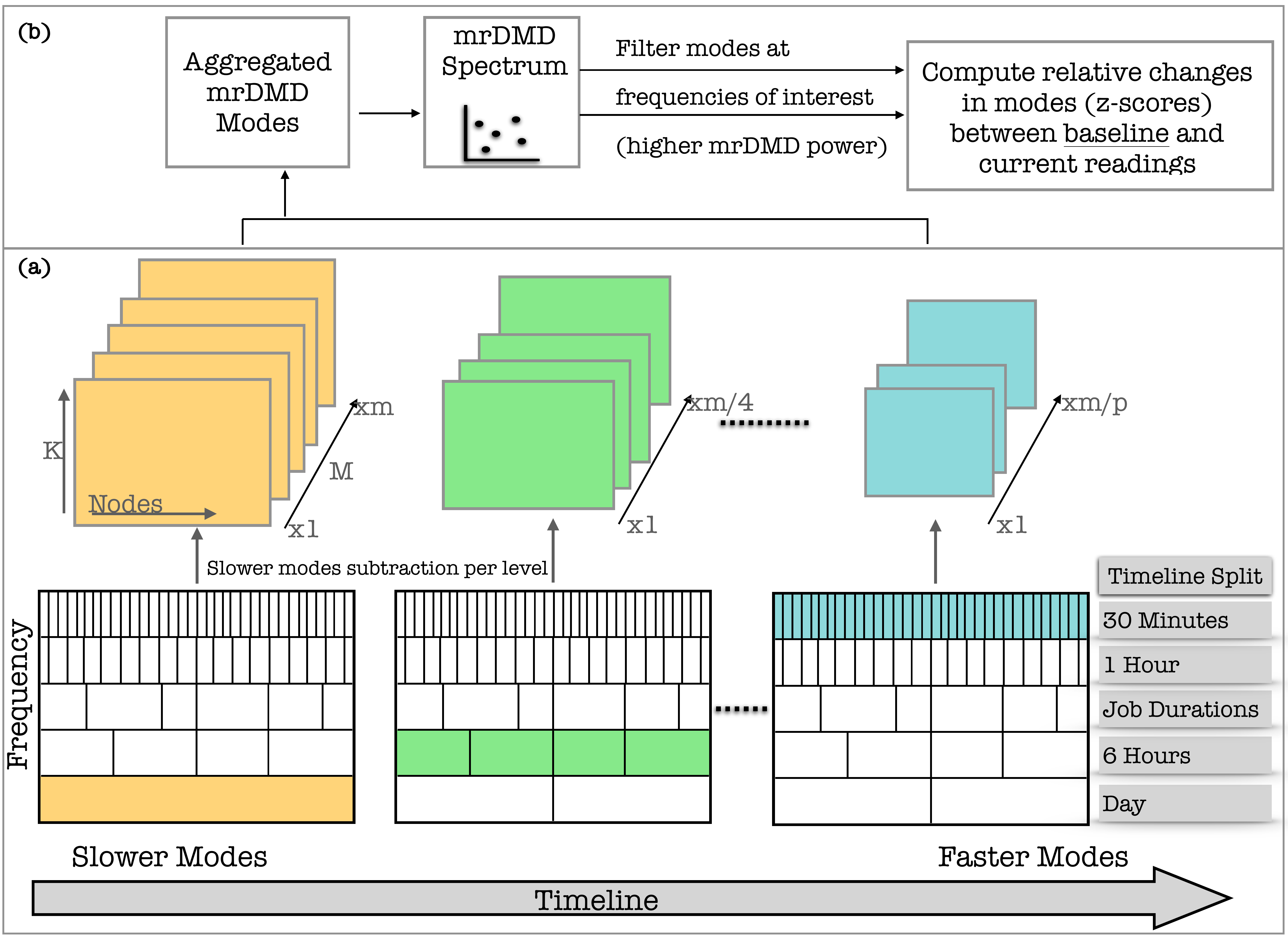}} 
\caption{Representation of the multiresolution dynamic mode decomposition~\cite{mrDMD}}
\label{mrdmd_method}
\end{figure}

We will now briefly introduce the DMD algorithm~\cite{tut, mrDMD, BRUNTON20161}.
Collecting $K$ sensor readings from $M$ snapshots in time, we may construct two raw data matrices $\mathrmbf{X}$ and $\mathrmbf{X}'$, in which $\mathrmbf{X}'$ columns are shifted by one time point from $\mathrmbf{X}$.

$
    \mathrmbf X = 
        \begin{bmatrix}
            | & | & | & & | \\ 
            \mathrmbf{x}_{1} & \mathrmbf{x}_{2} & \mathrmbf{x}_{3} & ... & \mathrmbf{x}_{M-1} \\
            | & | & | & & |
            
        \end{bmatrix}
$

$
    \mathrmbf X' = 
        \begin{bmatrix}
            | & | & | & & | \\ 
            \mathrmbf{x}_{2} & \mathrmbf{x}_{3} & \mathrmbf{x}_{4} & ... & \mathrmbf{x}_{M} \\
            | & | & | & & |
            
        \end{bmatrix}
$

The algorithm estimates the eigendecomposition for the best-fit matrix $\mathrmbf{A}$  such that

\begin{equation}
    \mathrmbf{X}' = \mathrmbf{AX}.
\end{equation}
The DMD algorithm has the following steps:\\
1) Decompose the data matrix $\mathrmbf{X}$ with SVD~\cite{numlin}:
    \begin{equation}
    \mathrmbf{X} = \mathrmbf{U}\boldsymbol{\Sigma} \mathrmbf{V^{*}}
    \end{equation}
where $*$ is the conjugate transpose, $\mathrmbf{U}\in \mathbb{C}^{K\times r}$ , $\boldsymbol{\Sigma} \in \mathbb{C}^{r\times r}$ and $\mathrmbf{V}\in \mathbb{C}^{M-1\times r}$. Here $r$ is the rank of the reduced SVD approximation to $\mathrmbf{X}$. The SVD rank reduction is computed using the optimal Singular Value Hard Threshold (SVHT~\cite{opsvht}).\\
2) Compute the $r\times r$ projection of the full matrix $\mathrmbf{A}$ onto $\mathrmbf{U}$'s low-rank modes, $\mathrmbf{A'}$.
\begin{equation}
\mathrmbf{A'} = \mathrmbf{U^{*}AU} = \mathrmbf{U^{*}X'V}\boldsymbol{\Sigma^{-1}}
\end{equation}
\\
3) Compute eigendecomposition of $\mathrmbf{A'}$ :
\begin{equation}
\mathrmbf{A' W = W}\boldsymbol{\Lambda}
\end{equation}
where eigenvectors are found in the columns of $\mathrmbf{W}$ and the corresponding eigenvalues $\mathrm{\lambda}_i$ are in diagonal matrix $\boldsymbol{\Lambda}$.\\
4) Compute the DMD modes
\begin{equation}
\boldsymbol{\Phi} = \mathrmbf{X'V}\boldsymbol{\Sigma^{-1}}\mathrmbf{W}
\end{equation}
Each column of $\boldsymbol{\Phi}$ contains a DMD mode $\mathrm{\phi_i}$ corresponding to the $ith$ eigenvalue.
Knowing the low-rank approximation of the data in the form of eigenvalues and eigenvectors, the projected future result can be constructed for all time and is given by
\begin{equation}
\mathrm{\tilde{x}(t) = \sum_{i=1}^{r} b_i(0)\phi_i(\boldsymbol{\xi}) exp(\omega_i t)} = \boldsymbol{\Phi}diag(exp(\omega t))\boldsymbol{b}
\label{eq:o}
\end{equation}
where $b_i(0)$ is the initial amplitude of each mode, $\boldsymbol{\Phi}$ is the matrix whose columns are the eigenvectors $\phi_i$, $\mathrm{\omega_i = ln(\lambda_i)/\Delta t}$, where $\Delta t$ is the time step, $diag(exp(\omega t))$ is a diagonal matrix whose entries are the eigenvalues $exp(\omega t)$, $\boldsymbol{\xi}$ are the spatial coordinates, and $\boldsymbol{b}$ is a vector of the coefficients $b_i$.

The mrDMD is a recursive algorithm that removes low-frequency content from snapshots of data collected over a period of time. Low-frequency content captures the slow-varying dynamics of the system. M is chosen so that it is large enough to represent the system's dynamics, i.e., enough high- and low-frequency components are present. At the initial recursion step, the slowest DMD modes (\textit{m1}) are subtracted from the DMD result (Fig.~\ref{mrdmd_method}(a)). The slower modes correspond to values below a threshold, computed by a user-defined maximum number of oscillations in the time series divided by the time range (i.e., the number of time points) at each split. We have chosen the maximum number of oscillations for our current dataset to be $4$. We then split the dataset into segments along the timeline (as shown in Fig.~\ref{mrdmd_method}(a)). DMD is once again separately performed on each split segment. The slowest modes are subtracted again, and the recursive algorithm continues until termination. In mrDMD, due to the subtraction of slower modes in the previous levels, at each level, the multiresolution features are represented by different spatiotemporal DMD modes. Therefore no single set of modes dominates the SVD and influences features at multiple levels or time scales.

The future solution for all time using the mrDMD low-rank approximation of the system as in Eq.~\ref{eq:o} is given by~\cite{mrDMD}:

\begin{equation}
\mathrm{x_{mrDMD}(t) = \sum_{i=1}^{M} b_i(0)\phi_i^{(1)}(\boldsymbol{\xi}) exp(\omega_i t)}
\end{equation}

\begin{equation}
\mathrm{= \sum_{i=1}^{m_1} b_i(0)\phi_i^{(1)}(\boldsymbol{\xi}) exp(\omega_i t) + 
\sum_{i=m_1+1}^{M} b_i(0)\phi_i^{(1)}(\boldsymbol{\xi}) exp(\omega_i t)} 
\label{eq:mr}
\end{equation}
 \indent \indent \indent \indent \textit{(slower modes)} $\qquad \qquad \qquad $\textit{(fast modes)} \\

 here the $\phi_i^{(1)}(x)$ represent the DMD modes computed from the full M snapshots. The first term in Eq.~\ref{eq:mr} has the slower dynamics whereas the second term has the faster dynamics. The second term gives the fast scale data matrix
 that is sent forward in the recursion step.

At each level of the data split, our algorithm automatically determines the sampling rate. Since the slower modes are subtracted at the initial levels, we use a lower sampling rate of the time series data at these levels. As the levels get higher, we increase the sampling rate, hence processing more time points (i.e., extracting higher frequency modes). For the supercomputer log, we determine the sampling rate as four times the Nyquist limit to capture cycles~\cite{nyq}. This value is customizable. At each level of the mrDMD analysis,  we asynchronously process each time range generated after the split, as the time range results within a level are independent of the other time ranges. This improvement reduces the computation time of the analysis (refer section~\ref{sec:Discussion} Fig.~\ref{perf}). 

The jobs utilizing the supercomputer execute for multiple hours. The original mrDMD algorithm split the timeline at each level into equal parts~\cite{mrDMD}. However, when handling environment data collected from multiple nodes in a supercomputer, a single timeline split may contain time series from multiple jobs. Since each job may utilize the system differently, mrDMD modes computed at each split may contain results from the time series of multiple jobs. This overlap of mrDMD modes will result in erroneous results, especially if the user wants to filter results per job. Therefore, we improved the algorithm by taking into account the start and end times of all jobs utilizing the nodes from which the time series are processed. We determine a value at which we split the timeline resulting in the minimum overlap between job start and end times. When analyzing time series data from multiple nodes, we calculate a time point for split where the distance between the start or the end times of  jobs in the vicinity ($\pm 2$ hours, this number depends on the level at which the job duration split (refer Fig.~\ref{mrdmd_method}) is performed could range from a few minutes to hours) of the chosen time point is minimum. This improves the final results, as seen in section~\ref{sec:CaseStudy} (Fig.~\ref{cs22}). We use job duration split at level 3 in our analysis (refer Fig.~\ref{mrdmd_method}(a))). However, it is customizable to be included in other levels depending on the length of the jobs and the initial length of the time series. 

\subsubsection{Frequency Isolation of spatiotemporal modes using mrDMD spectrum}

The mrDMD decomposition at each level results in spatial modes $\mathrm{\phi_i}$, each with a corresponding eigenvalue $\mathrm{\lambda_i}$ describing its temporal dynamics. The eigenvalue $\mathrm{\lambda_i}$ is  complex-valued. The dynamics can be easily interpreted after the above transformation $\mathrm{\omega_i = ln(\lambda_i)/\Delta t}$; the sign of the real component of $\mathrm{\omega_i}$ determines if the corresponding mode dynamics are growing (positive), or decaying (negative), and the imaginary component determines the frequency of oscillations. We compute the frequency of oscillation of mode $\mathrm{\phi_i}$ in units of cycles per second as 
\begin{equation}
\mathrm{f_i = \Bigg  | \frac{imag(\omega_i)}{2\pi}\Bigg |}
\end{equation}

Analogous to the traditional power spectrum computed with the fast Fourier transform (FFT) algorithm \cite{welch1967use}, the mrDMD ``power'' spectrum is then computed at each frequency of oscillation $\mathrm{f_i}$ and is given by $\mathrm{P_i = \big  | \phi_i\big |_2^2}$~\cite{BRUNTON20161}. Our mrDMD algorithm uses mrDMD mode magnitudes where the DMD spectrum power (normalized to 1.0) is greater than $0.5$. Our experiments have shown that this value helps adequately capture the low-rank representation of the dynamics of our current dataset. This improvement removes the noisy low power mode magnitudes and their corresponding frequencies and also helps reduce the computation time. 

 \begin{figure}[h]
	\centering
    \includegraphics[width=\linewidth]{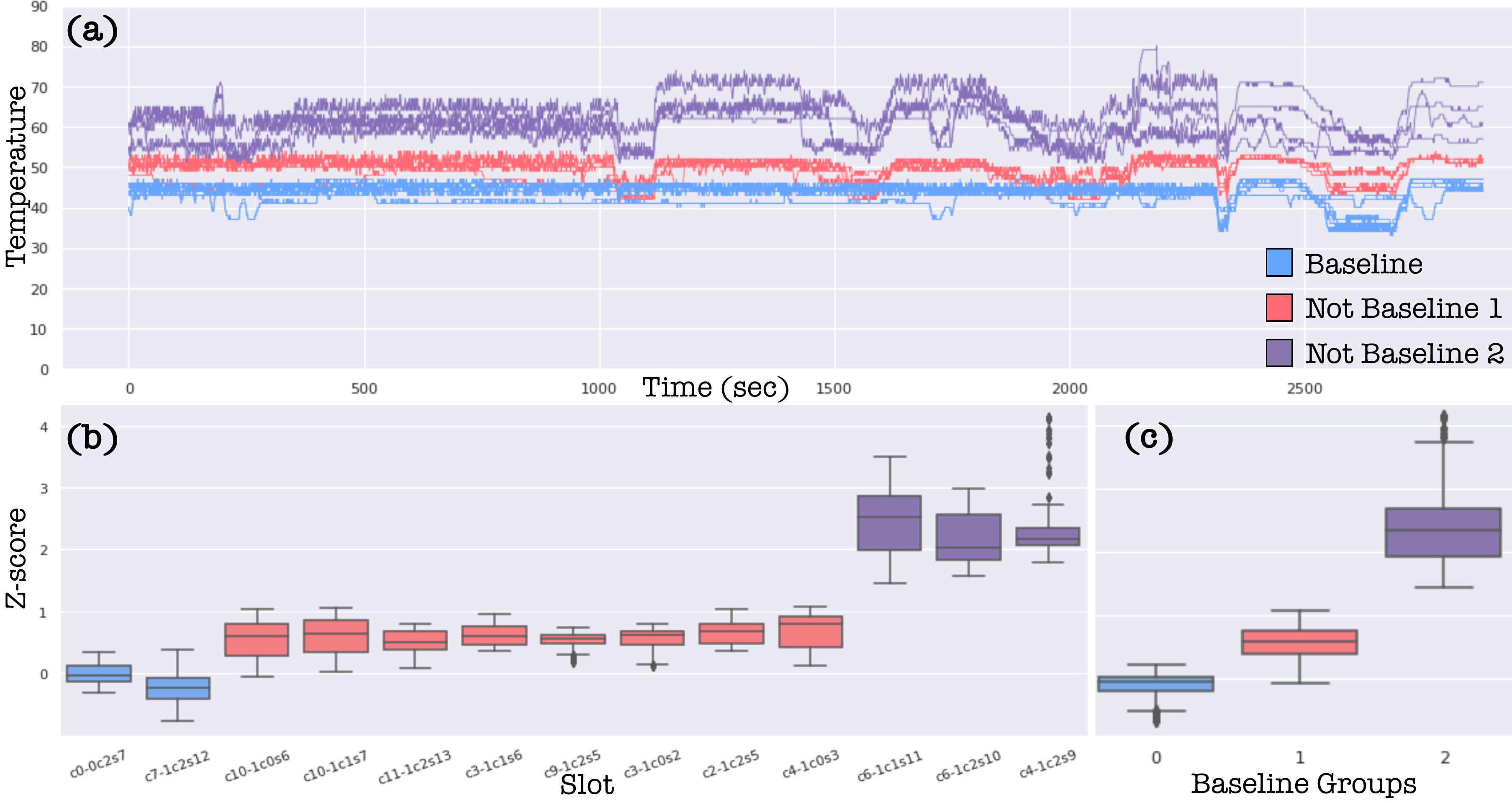}
    \caption{Analyzing (a) the supercomputer environment logs containing selected baseline readings(blue) and non-baseline readings(red, purple). (b) and (c) show the boxplots of the z-scores (change of non-baseline readings from baseline) versus slots(each slot contains 4 supercomputer nodes) or time series groups.}
	\label{baseline_method}
\end{figure}

\subsubsection{Baseline Selection}
Once we extract the high-power mrDMD modes from frequency ranges using the mrDMD spectrum, we compute the absolute values of the extracted mrDMD modes in the frequency ranges and average it to obtain the mean mode magnitudes. We do this separately for readings that we classify as baseline readings and non-baseline readings. Baseline readings constitute a subset of time series data that indicate an expected system functionality~\cite{BRUNTON20161}. We choose the baseline readings based on (1) the system specifications that constitute normal functioning, e.g., readings within the normal temperature range without causing the supercomputer to overheat, or (2) the user's usage trends as seen in the environment logs of the nodes used by the user's jobs. We use the system manuals to identify the baseline value ranges for the first baseline type. For the second baseline type, we pick the environment log readings from nodes allocated to a user's job that has completed a successful execution. 

We then compute the standard deviation of the non-baseline minus the baseline mean mode magnitudes at each reading (time series) for a trial set through bootstrapping. The trial set consists of time series following a typical scenario, e.g., passed job status, no hardware errors, etc. We then compute the z-scores of the difference from the baseline at each reading (time series) using the bootstrapped calculations of the standard deviation~\cite{BRUNTON20161}. Thus we obtain the relative changes in mean mode magnitudes for each new data using the previously computed results for data exhibiting normal behavior. For the system specification baseline, the z-scores capture the difference between what we identify as normal system (baseline) behavior and current system (non-baseline) behavior. For the user's usage trend baseline, the z-scores capture the difference between the utilization of the system by a user over time. 

Fig.~\ref{baseline_method} shows how baseline and non-baseline readings transform into z-score values. Fig.~\ref{baseline_method}(a) shows the temperature readings, grouped as baseline (blue) and non-baseline (purple and red) readings, for multiple nodes in the supercomputer. Fig.~\ref{baseline_method}(b-c) shows the z-scores versus slots and the reading groups (groups 0, 1, and 2 constitute blue, red, and purple readings, respectively). The reading groups are chosen based on the differences in their magnitudes and variations over time. Here as expected, the z-score values for baseline readings are close to $0$ (y-axis). However, for those readings away from the baseline, the readings with a smaller z-score change ($<1$) are in red and those with a larger change ($>1.5$) are purple. We further explain the use of these types of baselines in the case study section~\ref{sec:CaseStudy}.

\subsection{Visualization Frontend}

\begin{figure}[h]
	\centering
    \includegraphics[width=0.7\linewidth]{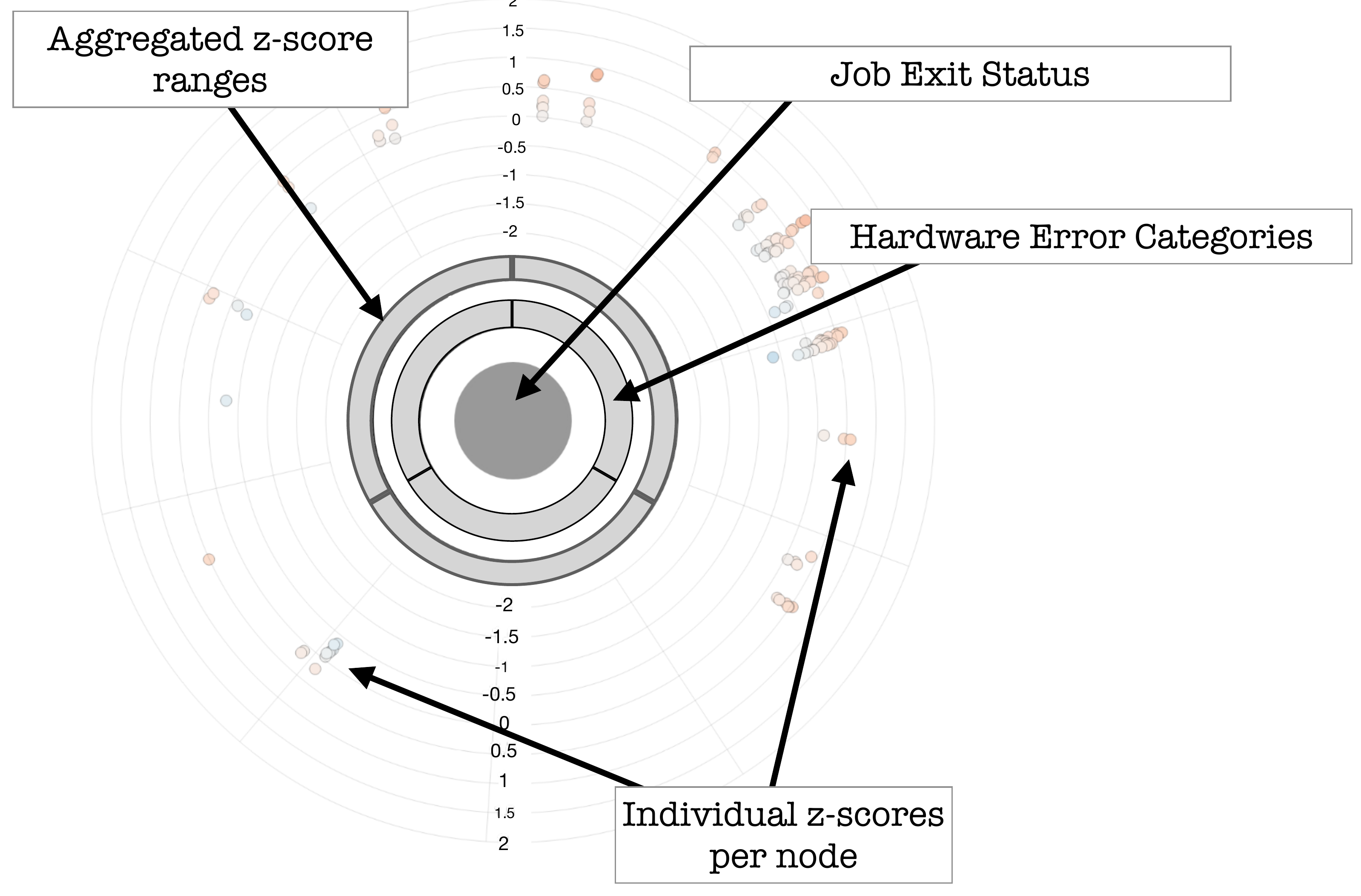}
    \caption{Abstract and glyph view layouts}
	\label{glyphv}
\end{figure}

\begin{figure*}[t]
\centerline{\includegraphics[width=1\textwidth]{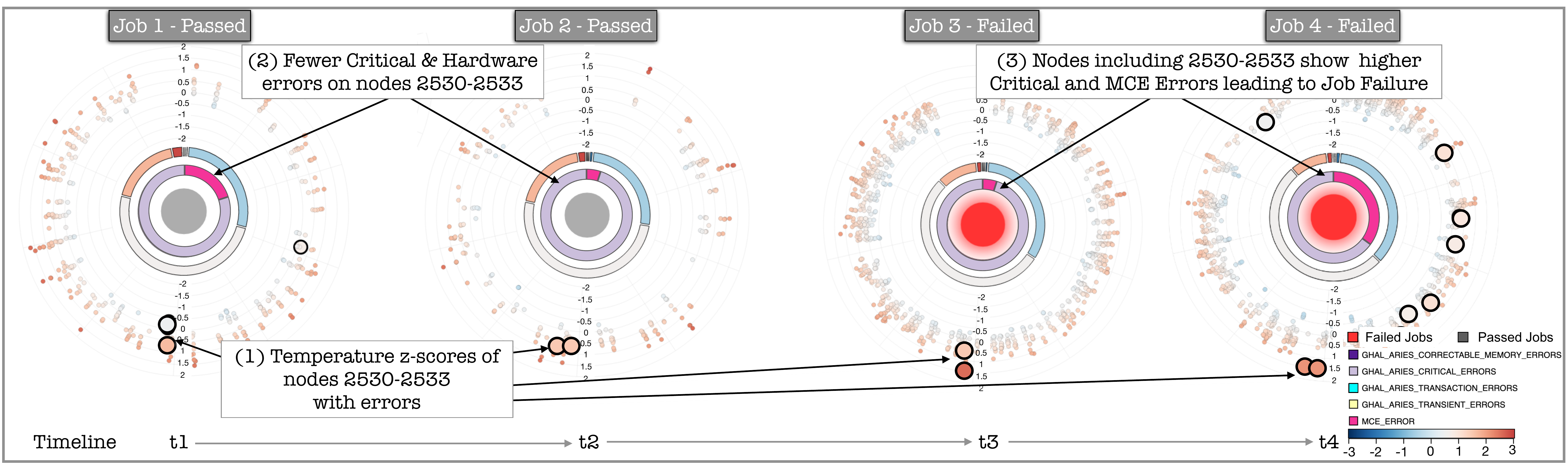}} 
\caption{Glyph view of four jobs utilizing the supercomputer. The job glyphs are sorted in the ascending order of their start times. The first two jobs ($1$ and $2$) have exited with a pass status and the last two ($3$ and $4$) have failed. Nodes $2530-2533$ consistently report hardware (MCE and critical) errors. The z-score values, computed using the system specification baseline, increased for the nodes $2530-2533$ from job $1$ to job $4$ resulting in node $2533$ shutdown along with job $4$ failure.}
\label{cs1_1}
\end{figure*}

\begin{figure}[h]
\centerline{\includegraphics[width=\linewidth]{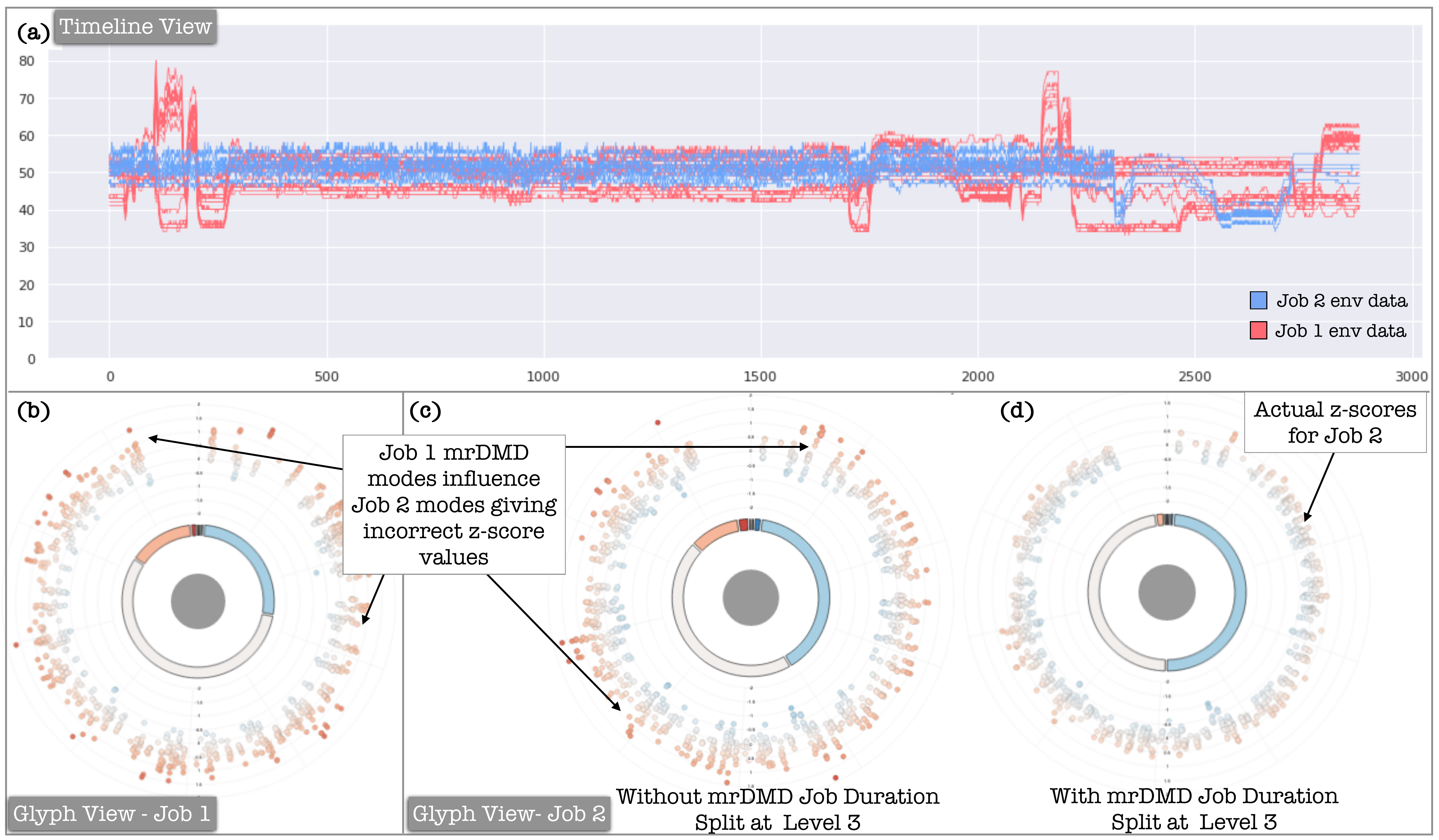}} 
\caption{Analysis of two jobs utilizing the system in the same spatial and temporal locality. (a) shows a subset of the environment log readings for job 1 (red) and job 2 (blue), (b) shows job 1 glyph view, and (c)(d) shows glyph views for job 2 without and with the mrDMD timeline split for job duration at level 3, respectively.}
\label{cs22}
\end{figure}

Fig.~\ref{ui} shows the UI of our visual analytics tool. The abstract view (Fig.~\ref{ui}(a)) shows the aggregated values of the log data.
 The glyph view (Fig.~\ref{ui}(b)) shows the log data
information separated by granularities of jobs, users, projects, nodes used, and job exit codes. The glyph view uses the force directed layout algorithm~\cite{10.5555/2386124.2386138}. The abstract and the glyph view sections (Fig.~\ref{glyphv} and Fig.~\ref{ui}(a),(b)) contain a center circle specifying if the job has passed (gray) or failed (red). The innermost pie chart shows the hardware log error categories grouped by their counts for the job run duration. The glyph view (Fig.~\ref{ui}(b) and Fig.~\ref{glyphv}) has another concentric pie chart that shows the z-score values grouped by counts within one standard deviation range (which is customizable) from values $-2$ to $2$.  Note that z-score values can range from $-10$ to $10$ in the supercomputer log dataset.
 The z-scores have a lighter white hue if they lie closer to the baseline, i.e., 0. We display the counts of nodes on the mouseover interaction over each arc of this pie chart. The outermost radial view shows the scatter plot of the z-scores for each node utilized by the job. Since there are $4,300$ nodes, the radial axis takes $4,300$ node numbers (IDs) in clockwise order. The node history view (Fig.~\ref{ui}(c)) shows the history of the nodes selected by lasso selection in the abstract and the glyph views (Fig.~\ref{ui}(a,b)). In the timeline view (Fig.~\ref{ui}(d)), we display environment log data (power, temperature, voltage, current, etc.). The timeline view has a built-in brush interaction using which one can refine the results of the node history view (Fig.~\ref{ui}(c)). The job view uses parallel coordinates to display the details of the job log information. On clicking each job in this view, the glyph and node history view results get updated. As a part of our future work, we plan to minimize the z-score overlap in the radial view through aggregations. Our current implementation provides scrolling interaction to zoom in to a section of the radial axis, thus eliminating the overlaps.

In Fig.~\ref{ui}(a), the user has selected nodes from the z-score radial view. The size of the nodes is proportional to the size of the fatal errors in the system for the time duration specified in the timeline view (Fig.~\ref{ui}(d)). This example shows an 11-hour duration data. Using the lasso selection, the user can select these nodes to view them in finer detail. On node selections, the glyph view (Fig.~\ref{ui}(b)) gets updated, and here it shows the jobs that have been executed in this duration of our analysis. In this view, we can choose other granularities such as jobs, users, projects, nodes used, and job exit codes. The job details are plotted in the parallel coordinates view (Fig.~\ref{ui}(e)), giving more information about its specifications. The node history view (Fig.~\ref{ui}(c)) then shows the current errors and history of the errors in the selected nodes and the jobs that encountered these errors. Here Nodes $0$ and $3$ in slots $c0-0c1s12$ are the problematic nodes reporting transaction and MCE (machine check exception) errors.

\begin{figure*}[t]
\centerline{\includegraphics[width=1\textwidth]{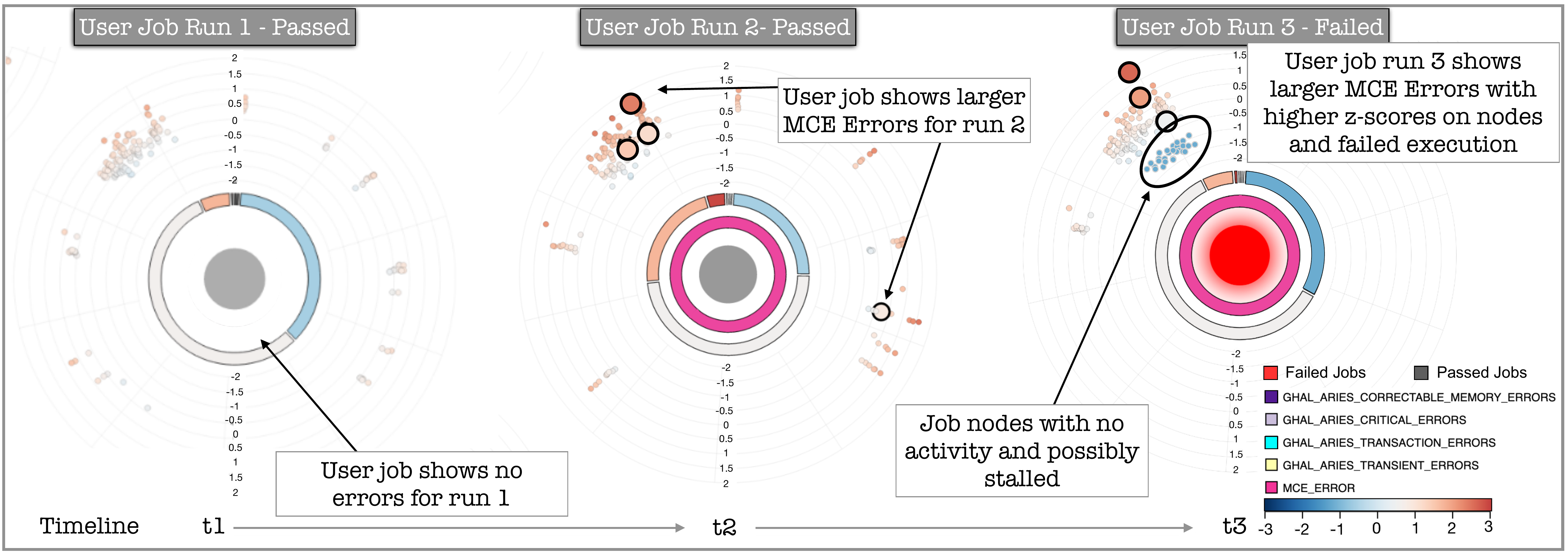}} 
\caption{Glyph view of 3 jobs utilizing the supercomputer. The job glyphs are sorted in the ascending order of their start times. The first two jobs ($1$ and $2$) have exited with a pass status and the last one ($3$) has failed. Few nodes consistently report hardware (MCE) errors. The z-score values, computed using user specific baseline, increased for the error nodes in jobs $2$ to job $3$. Job $3$ showed a subset of nodes that were underutilized and were possibly stalled.}
\label{cs2}
\end{figure*}

\begin{figure*}[t]
\centerline{\includegraphics[width=1\textwidth]{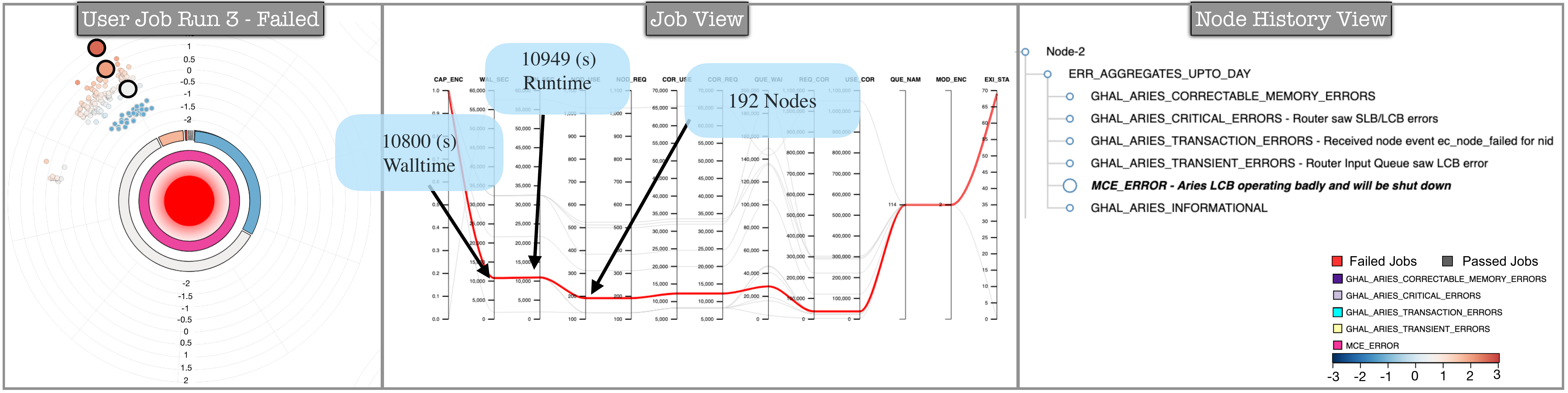}} 
\caption{Analysis of job 3 from Fig.~\ref{cs2} showed that the job exceeded the allocated time and consistently reported errors in the duration of its runtime with possibly underutilized nodes.}
\label{cs2_2}
\end{figure*}

\section{Case Studies}
\label{sec:CaseStudy}

\subsection{Case Study 1}
We describe two detailed cases of supercomputer utilization by user jobs and present how our system aids in identifying usage patterns from
large-scale and diverse logs. We have shown an analysis example in Sec. 2.3. The case studies further demonstrate the mrDMD results using our visual analytics tool and compare the processed data from each log type mentioned previously. 
Our environment logs are collected from the Cray XC40 supercomputer and contain data from $24$ compute racks with a total of $4,300$ nodes. Sensors on each node collect data in $10$-$30$ second intervals. We mainly analyze the temperature and current readings from the environment logs. There are 4 temperature and 4 current readings per node. We perform the mrDMD analysis and compute the z-scores separately for each type of reading.

\begin{figure*}[b]
	\centering
    \includegraphics[width=\linewidth]{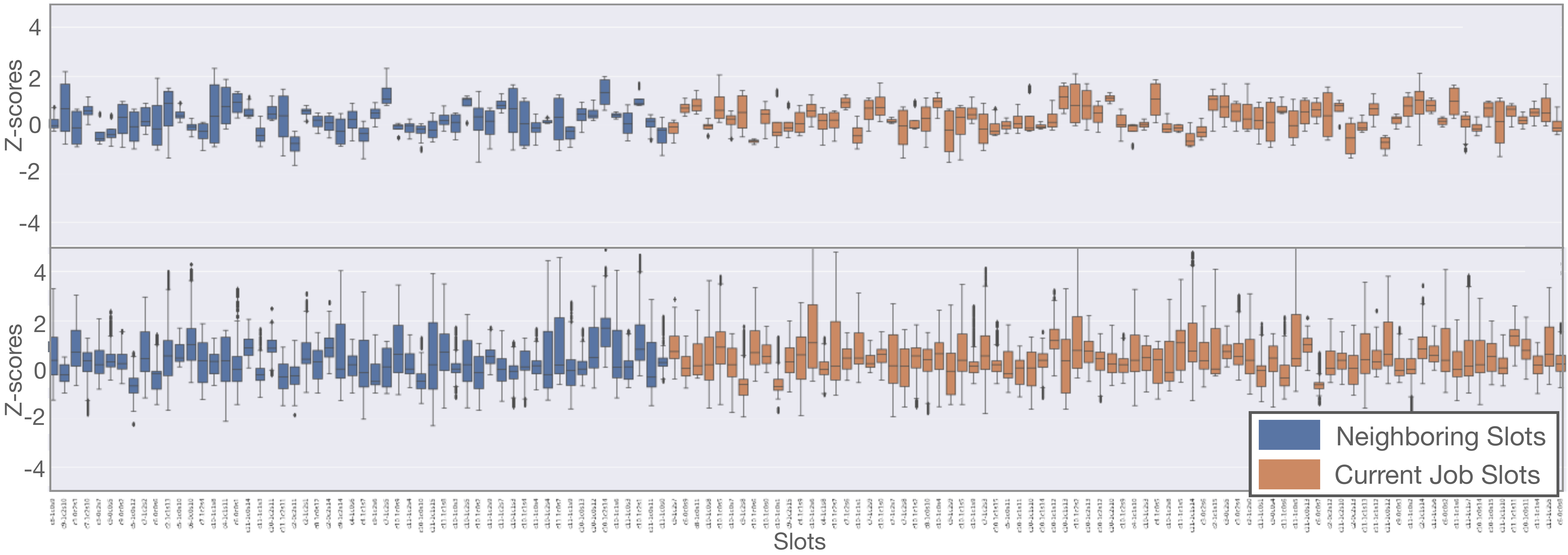}
    \caption{Analyzing the supercomputer log data showing z-score values versus nodes for one job run. The top (a) and bottom (b) figures show the z-scores for a subset of job node readings and neighboring node readings before and after the job run reported hardware errors, respectively. The slots along the x-axis are sorted by their node IDs}
	\label{dis3}
\end{figure*}

The Fig.~\ref{cs1_1} shows the glyph view for four different jobs utilizing the system starting from time $t1$ to $t4$. Here we ordered the job glyphs in the ascending order of their start time, i.e., $t1<t2<t3<t4$. Each glyph contains a center circle that specifies if the job has passed or failed. The first two jobs have passed (gray), and the last two have exited with a failed (red) status. The innermost pie chart shows the hardware log error categories grouped by their counts for the job run duration. Fig.~\ref{cs1_1} shows that the four jobs have both MCE (machine check exceptions) and critical errors throughout the job runs. The next concentric pie chart shows the z-score values grouped by counts within one count range from values $-2$ to $2$. Note that z-score values can range from -10 to 10 in the supercomputer log dataset, but they vary from $-2$ to $2$ z-score range for the current temperature examples. The z-scores have a lighter white hue if they lie closer to the baseline, i.e., 0. Here we see that most of the z-scores follow typical behavior and hence have a lighter hue. We display the counts of nodes within each z-score range on the mouseover interaction on each arc of the pie chart. The outermost radial view shows the scatter plot of the z-scores for each node utilized by the job. Since there are $4,300$ nodes, the radial axis takes $4,300$ node numbers (IDs) in clockwise order. The larger nodes are reporting hardware errors. We have the hover interaction, which gives the node IDs on the mouseover interaction. The user can also pan or zoom in to a region of the radial axis, which helps a user scan the visualization in more detail. The four jobs chosen ran during a 24-hour duration and utilized a set of nodes with IDs 2430-2533 that consistently reported hardware errors. We have used system specification baselines for all four jobs.
We see from job 1 that the z-score move from being closer to the baseline (white, light blue hues) to being much higher than the baseline (orange hues) for job 4. Orange hues signify higher temperature on the nodes, and this temperature increase will affect the neighboring nodes as well (see Fig.~\ref{dis3}). Also, the number of nodes reporting hardware errors increased significantly at job 4, resulting in a node failure of node 2533 and eventually the job 4 failure. Finally, the node history view (not shown in Fig.~\ref{cs1_1}) summarized additional details of the errors as \textit{Router input queue saw SLB error},  \textit{Aries LCB operating badly and will be shut down}, and \textit{Node ID 2533 unavailable}. Our tool helps visualize these consistent patterns of failures for the first three job runs, indicating that these anomalous nodes need to be checked by system administrators and any hardware failures be immediately rectified before a node shuts down, as in this case. Another usefulness of utilizing z-scores from mrDMD analysis of environment logs is that nodes in the immediate vicinity also report similar hardware errors, although this may not always lead to node shutdown. 

Fig.~\ref{cs22} shows the analysis of two jobs utilizing the system in the same spatial locality. Job 1 had executed before job 2 (Note that we have overlayed 3000 time points of both jobs along the x-axis for clarity). Fig.~\ref{cs22}(a) shows a subset of the temperature readings reported by the nodes utilized by the jobs, job 1 (red) and job 2 (blue). Temperature readings for job 2 mostly follow close to the selected baseline with a range between $40^\circ$- $60^\circ$ Celsius. However, the readings greatly fluctuate for job 1, between $30^\circ$- $80^\circ$ Celsius.
Fig.~\ref{cs22}(b) shows the glyph view for job 1, as expected, the z-scores are largely varying as it follows the time series pattern shown in Fig.~\ref{cs22}(a). Fig.~\ref{cs22}(c)(d) show the glyph views for job 2 without and with the mrDMD timeline split for job duration at level 3, respectively. Without the job duration split in the mrDMD analysis (Fig.~\ref{cs22}(c)), the z-score patterns for job 1 and job 2 look similar even though that behavior is not reflected Fig.~\ref{cs22}(a). The mrDMD analysis could give erroneous results due to overlapping time ranges between the end of job 1 and the beginning of job 2. This results in higher frequency modes between jobs to overlap and, in some cases, leads to erroneous results, as seen in Fig.~\ref{cs22}.

\subsection{Case Study 2}

We utilize environment logs from the Cray XC40 supercomputer that contain data from 24 compute racks with a total of 4,300 nodes. In this case study, we filter jobs by a single user. The figure shows the glyph view for three different jobs from a single user utilizing the system starting from time $t1$ to $t3$. Here we ordered the job glyphs in the ascending order of their start time, i.e., $t1<t2<t3$. Each glyph contains a center circle that specifies if the job has passed or failed. The first two jobs have passed (gray), and the last one has exited with a failed (red) status. The innermost pie chart shows the hardware log error categories grouped by their counts for the job run duration. Fig.~\ref{cs2} shows that two out of three jobs have predominantly reported MCE (machine check exceptions) errors throughout the job runs. Note that we have filtered out the informational type hardware log data. The next concentric pie chart shows the z-score values grouped by counts within one count range from values $-2$ to $2$. The larger nodes are reporting hardware errors. The user can also pan or zoom in to a region of the radial axis, which helps a user scan the visualization in more detail and controls visual clutter. The three user jobs chosen ran during a 24-hour duration and utilized $192$ nodes. User jobs $2$ and $3$ consistently reported a larger number of hardware MCE errors on three nodes.

\label{sec:Discussion}

In this case study, for the mrDMD analysis, we chose the user-specific baseline readings. We picked the baseline readings from user job 1. The user job 1 (Fig.~\ref{cs2}) utilized 192 nodes with two readings per node, giving a total of $384$ readings. Since we select baselines from user job $1$ (one reading per node), the z-scores are mostly within the range ($-1$ to $1$), i.e., they lie closer to the baseline value $0$, with roughly one standard deviation variation on either side. However, for user jobs $2$ and $3$, the z-scores lie from $-2$  to $2$. A reason could be that the user made improvements or changes to their code to utilize the system at varying capacities. In this case, user job $2$ showed significant MCE errors in $4$ nodes and increased z-score values on one node ID $4505$. The next run, user job $3$, showed MCE errors with much higher z-score values, particularly on one node ID $4507$. However, we also see a small subset of nodes having a z-score much lesser than the baseline. This indicates that the nodes are not being utilized, as lower z-score values indicate a lesser current on these nodes compared to the previous runs. Also, low values of z-scores indicate a possible stalling where nodes are waiting without performing any computation for the duration of the job run. 

In Fig.~\ref{cs2_2}, we analyze job $3$ further using the auxiliary views, i.e., the job view and the node history view. On selecting the job $3$ in the glyph view, the job is highlighted in the job view, and we see in Fig.~\ref{cs2_2} that the runtime, i.e., the execution time of the job had exceeded the wall time (values visible through hover interaction), i.e., the maximum time allocated to the job. This scenario occurs when the application has stopped, but the job is still perceived to be running by the system. In such cases, the system resources could be in a stalled state and hence are unused, bringing down the usability of the overall supercomputer. Using this insight, the system admins could request a system recovery and repair in case of a consistent system errors across jobs. Furthermore, the system admins could inform the user to either fully utilize the allocated nodes or request fewer nodes on subsequent job runs. In this case study, the node history layout (Fig.~\ref{cs2_2}) indicates consistent \textit{MCE} and \textit{Aries LCB errors on the network link}. However, the node history also revealed memory-related, critical, transient, and transaction errors when utilized by other jobs in the past $24$ hours.

\section{Discussion}
\begin{figure}[h]
	\centering
    \includegraphics[width=\linewidth]{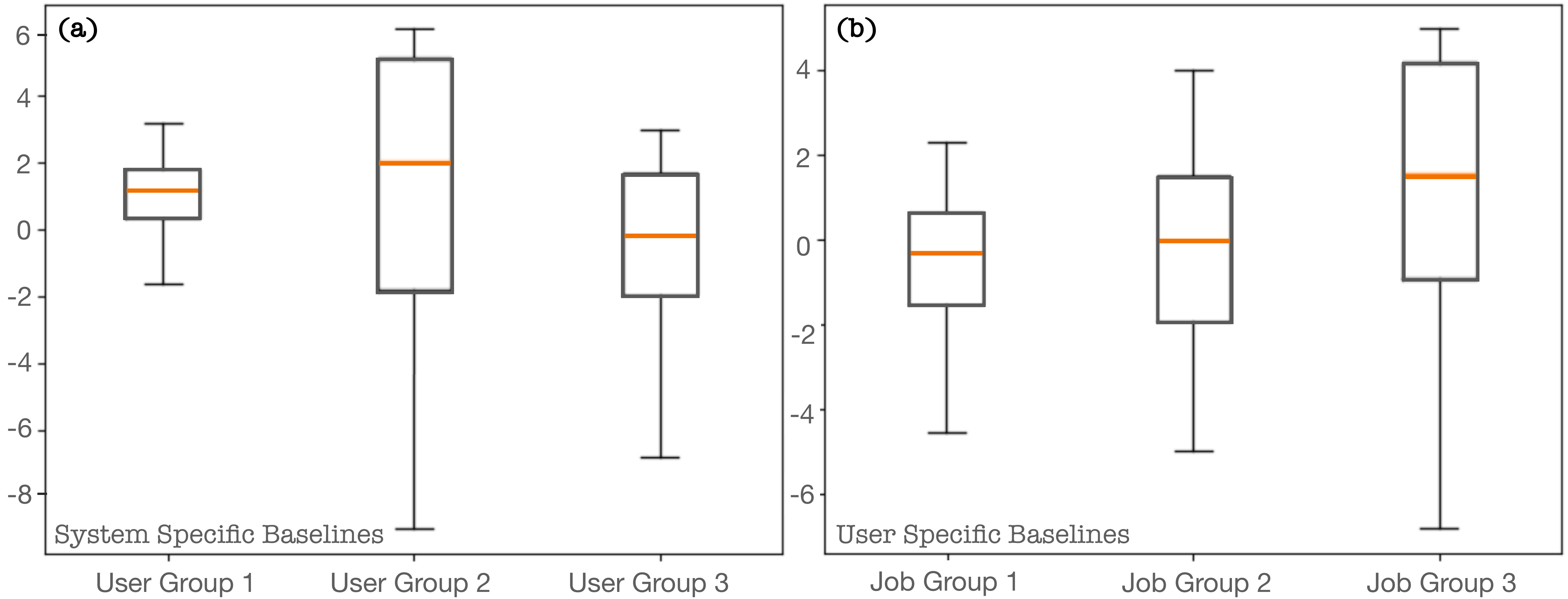}
    \caption{Analyzing the supercomputer log data using the (a) system-specific baseline and (b) user-specific baseline.}
	\label{dis1}
\end{figure}

We have presented our approach's effectiveness in analyzing large-scale time series data and its relevance to processing supercomputer environment logs to draw meaningful insights regarding the system's functionality. We further discuss the strengths of our approach in this section in categorizing jobs and users of the supercomputers using temperature readings. 

\begin{figure}[h]
	\centering
    \includegraphics[width=\linewidth]{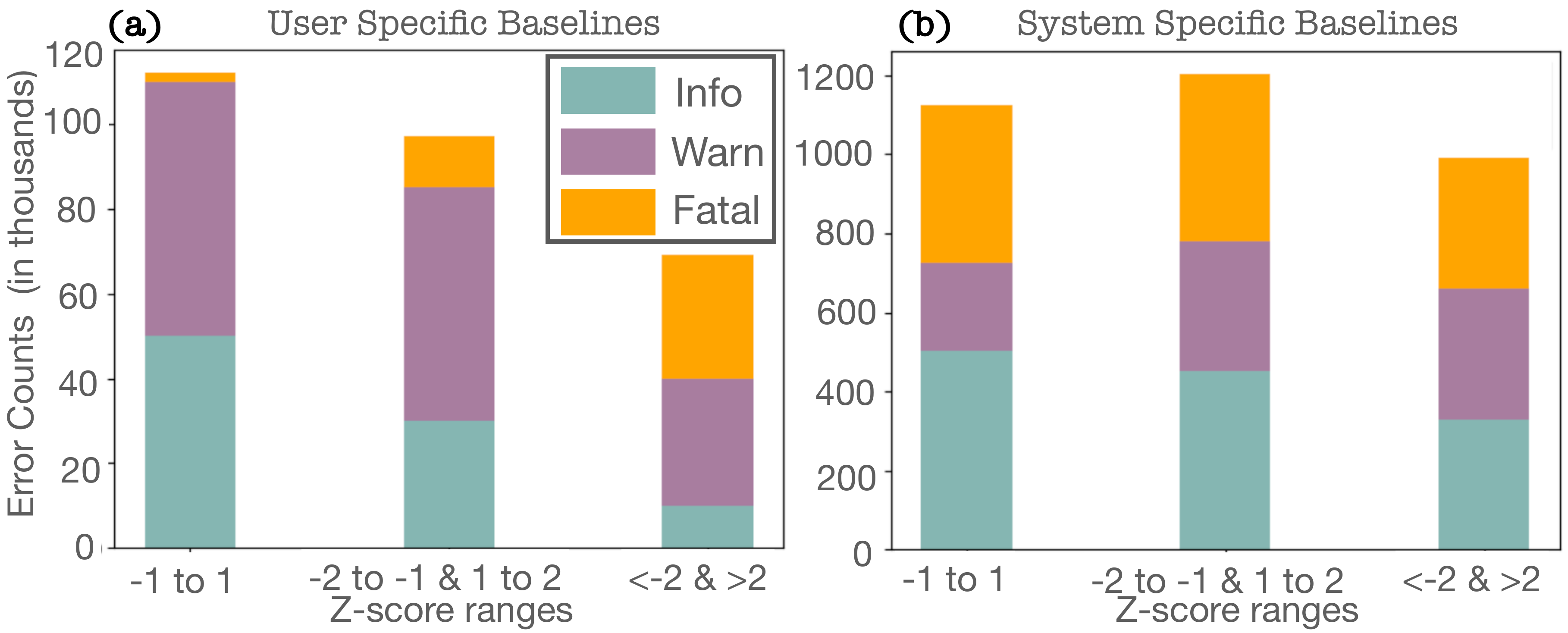}
    \caption{Analyzing the supercomputer log data using z-score ranges and plotting the counts of hardware errors at each range for (a) user-specific baselines and (b) system-specific baselines. (a) and (b) show the z-scores values before and
after the system reported hardware errors, respectively.}
	\label{dis2}
\end{figure}

First, we applied our mrDMD analysis to environment logs. We store the z-score results in an Elasticsearch database. We then isolated 11,654 jobs spanning the years 2019-2020. Fig.~\ref{dis1}(a) shows the boxplot of z-scores versus user groups for the supercomputer. Using the system-specific baseline, we were able to isolate particular groups of user utilization patterns on the supercomputer using temperature readings. The x-axis of Fig.~\ref{dis1}(a) shows users' job usage patterns that are similar, i.e., there is no overlap of users along the x-axis groups. User group 1 mainly consists of small runtime jobs ($<2$ hours) with z-scores slightly over the baseline. User group 2 consists of larger jobs, usually utilizing more than 256 nodes in the system. Most of the user group 2 jobs are high-capability jobs that fully use the memory and show higher temperature readings at most nodes throughout the job runtime, which was greater than 4 hours. User group 3 also consists of longer running jobs ($>2$ hours). However, unlike user group 2, these jobs are not high-capability. Knowing this information about supercomputer usage patterns, the system admins can encourage the users to request fewer nodes when they do not fully utilize their allocated nodes for their jobs. For shorter-running jobs, the system admins can reduce the user's wait times for node allocation and hence improve their job scheduling strategies~\cite{9355229}. 

Fig.~\ref{dis1}(b) shows the boxplot of z-scores versus job groups for the supercomputer. Using the user-specific baseline, we were able to isolate particular groups of job utilization patterns on the supercomputer using temperature readings. The x-axis of Fig.~\ref{dis1}(b) shows job patterns that are similar, i.e., there could be an overlap of users along the x-axis groups. Here, for each user, we chose one job that had no hardware errors and executed with a pass status. We then used the standard deviation computed from the mrDMD mode magnitudes of the baselines for the users' subsequent job runs. We repeated this process for over 300 users utilizing the supercomputer. Fig.~\ref{dis1}(b) shows the general categorization of user jobs when using user-specific baselines per user. The job group 1 consists of jobs with a lower number of hardware errors, and the environment logs follow the same z-score trend as the baseline jobs. The job group 2 and 3 both show a larger number of hardware errors. However, job group 3 consists mainly of high-capability jobs when compared to job group 2.  
From Fig.~\ref{dis1}(a) and (b), using the mrDMD analysis, we can categorize users and jobs using z-score values, and the system admins can track the system usage patterns compared to past utilization trends.  

Fig.~\ref{dis2} shows the plots for the hardware error counts versus z-score ranges. Here the hardware errors in the dataset are categorized into informational, warning, and fatal (errors linked with system failures) type error messages. Our goal was to check if the environment log information, displayed through z-scores, captured any hardware error trends. Fig.~\ref{dis2}(a) shows the plot for a group of 150 jobs for 23 users in the system. We compute the z-scores using a user-specific baseline for passed job runs. Here we see that the fatal hardware error counts increase as the z-score values move away from the baseline. However, the informational and warning messages are higher in lower z-score ranges. Fig.~\ref{dis2}(b) uses system-specific baselines to compute z-scores for 500 user jobs. However, when we attempted to find patterns of z-scores linking hardware error categories to specific groups of users or jobs, we could not derive a link. Fig.~\ref{dis2}(b) shows that the hardware error categories are roughly equal across all z-score values when using system-specific baselines for all user jobs. This demonstrates the usefulness of using user-specific baselines as they help identify and link hardware errors with user jobs.

Fig.~\ref{dis3} shows the plots of z-score values versus slots ($4$ nodes) in the supercomputer for one job run. The orange-colored slots are allocated to this job, and the blue belong to the neighboring slots not utilized by the job. Fig.~\ref{dis3}(a) and Fig.~\ref{dis3}(b) show the z-scores values before and after the system reported hardware errors, respectively. We can clearly see that z-scores show larger variations ($-4$ to $4$) in values after the job slots reported hardware errors. These hardware errors were memory and critical errors, and the resulting temperature spike affected not only the job slots but also the neighboring slots that reported no errors. Similar trends are visible across multiple jobs and further analysis is needed to identify countermeasures to avoid system heating or overheating. 

\begin{figure}[!h]
	\centering
    \includegraphics[width=\linewidth]{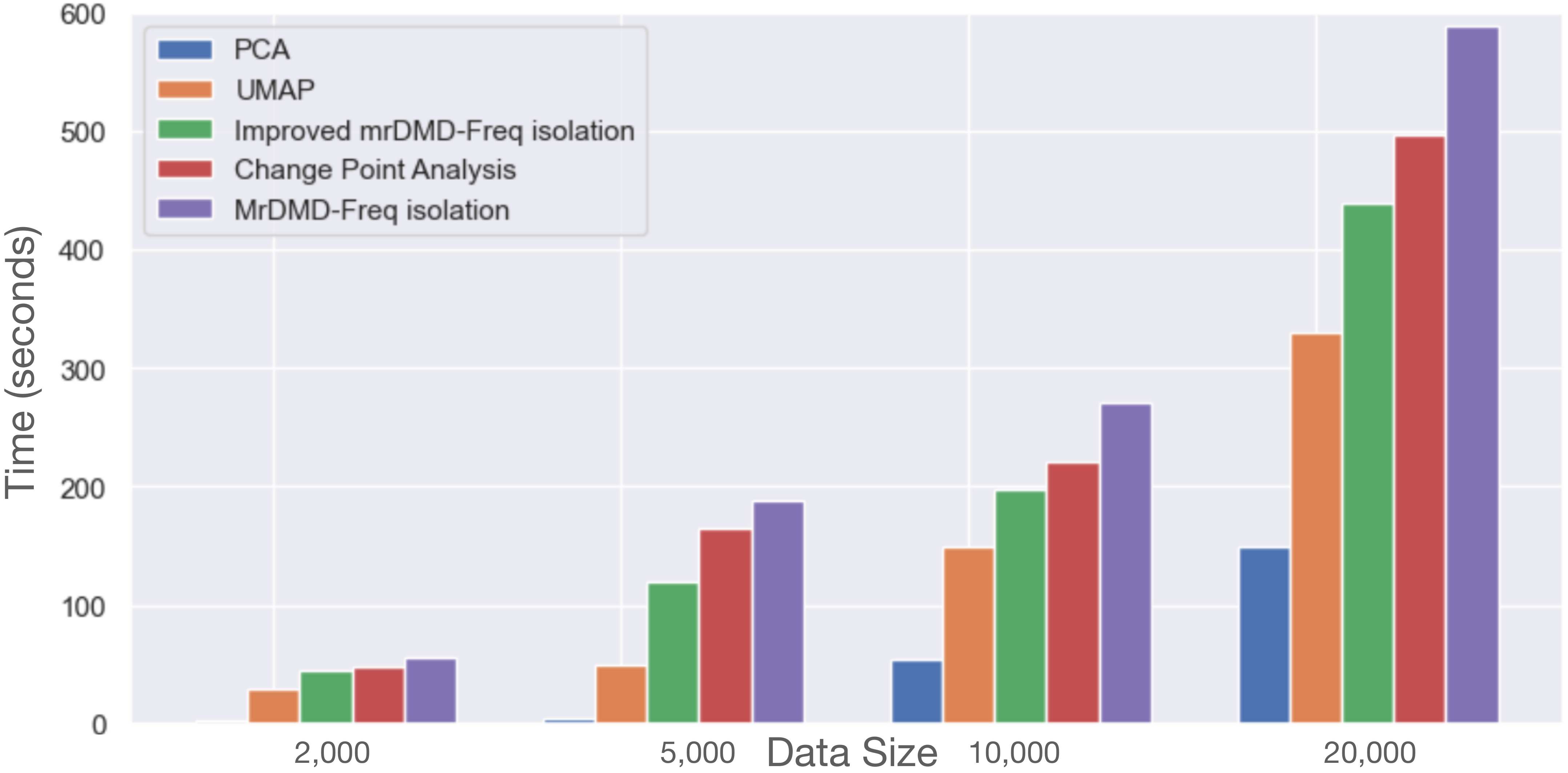}
    \caption{A comparison of completion times, showing how performance scales with the data size for $4,300$ readings.}
	\label{perf}
\end{figure}

The sampling rate at each level of our mrDMD depends on the level. Lower levels use lower sampling rates, and higher frequency levels sample data at a higher frequency. Since we do not extract high-frequency modes at lower levels, we can use this improvement without changing the final result. At each level of the mrDMD analysis, the algorithm subtracts the slower modes and splits the timeline. We process each of these splits asynchronously during the recursive process, adding a performance improvement, as shown in Fig.~\ref{perf}. For this analysis, we have picked some popular time series analysis and clustering methods, including PCA~\cite{Jolliffe2011}, UMAP~\cite{mcinnes2018umap}($\mathrm{n\_neighbors}\!=\!4$, $\mathrm{min\_dist}\!=\!0.1$, and $\mathrm{metric}$=``euclidean''), change point analysis~\cite{TRUONG2020107299}($\mathrm{model}$=``rbf'', $\mathrm{jump}$=3, $\mathrm{min\_size}$=2, $\mathrm{breakpoints}$=20). We then compare the computation times using $4,300$ time series of varying length data size, including $4,300\times2000$, $4,300\times5000$, $4,300\times10000$, and $4,300\times20,000$. We performed the experiments on the Cray XC40 supercomputer with two 6-core, 2.4-GHz Intel E5–2620 v3 processors (Intel Haswell architecture) and $256$GB of DDR4 memory. The results show that although our method does not outperform PCA and UMAP, it does outperform the change point detection algorithm and the mrDMD algorithm without our performance improvements.

\section{Conclusion}
\label{sec:Conclusion}

With significant achievements in computer engineering and the advent of exascale systems, there is a significant increase in monitoring data collected at multiple fidelity levels and varying temporal resolutions.
Our work aims to build a holistic visual analytical tool that processes this massive monitoring data, mainly the hardware logs, job logs, and environment logs collected from disparate subsystems and components of a supercomputer system. The tool provides an abstract overview of the systems underlying state and behaviors.
The main goal of our work is to consolidate and provide analysis results from multiple types of log data. Environment log data is usually ignored in log analysis mainly because of its massive size and lack of a mechanism to draw interesting patterns promptly. Our work aims to bridge this gap through our modified mrDMD analysis and visual analytics tool. With our improvements to the multiresolution dynamic mode decomposition (mrDMD) algorithm, we are able to promptly extract supercomputer usage and error patterns at varying temporal and spatial resolutions. We identified two types of baselines, representing a normal system behavior, and used them to extract groups of users and jobs that follow similar supercomputer usage patterns. Our data-driven analysis can be easily applied to other large-scale system log data. We plan to extend our work to support the analysis of streaming multivariate time series to deliver results in real time.

\section*{Acknowledgment}
This research was funded in part and used resources of the Argonne Leadership Computing Facility (ALCF), a DOE Office of Science User Facility supported under Contract DE-AC02-06CH11357. We thank the ALCF HPC Systems and Network Administration team, including George Brown, Ben Lenard, Eric Pershey, Ryan Milner, Doug Waldron and Jon Bouvet from the Cray team, for their guidance in comprehending various subsystems and features. 

\bibliographystyle{IEEEtran}
\bibliography{00_main}

\begin{thebibliography}{10}
\providecommand{\url}[1]{#1}
\csname url@samestyle\endcsname
\providecommand{\newblock}{\relax}
\providecommand{\bibinfo}[2]{#2}
\providecommand{\BIBentrySTDinterwordspacing}{\spaceskip=0pt\relax}
\providecommand{\BIBentryALTinterwordstretchfactor}{4}
\providecommand{\BIBentryALTinterwordspacing}{\spaceskip=\fontdimen2\font plus
\BIBentryALTinterwordstretchfactor\fontdimen3\font minus
  \fontdimen4\font\relax}
\providecommand{\BIBforeignlanguage}[2]{{%
\expandafter\ifx\csname l@#1\endcsname\relax
\typeout{** WARNING: IEEEtran.bst: No hyphenation pattern has been}%
\typeout{** loaded for the language `#1'. Using the pattern for}%
\typeout{** the default language instead.}%
\else
\language=\csname l@#1\endcsname
\fi
#2}}
\providecommand{\BIBdecl}{\relax}
\BIBdecl

\bibitem{mrDMD}
\BIBentryALTinterwordspacing
J.~N. Kutz, X.~Fu, and S.~L. Brunton, ``Multiresolution dynamic mode
  decomposition,'' \emph{SIAM Journal on Applied Dynamical Systems}, vol.~15,
  no.~2, pp. 713--735, 2016. [Online]. Available:
  \url{https://doi.org/10.1137/15M1023543}
\BIBentrySTDinterwordspacing

\bibitem{dmd_2009}
C.~W. ROWLEY, I.~MEZIĆ, S.~BAGHERI, P.~SCHLATTER, and D.~S. HENNINGSON,
  ``Spectral analysis of nonlinear flows,'' \emph{Journal of Fluid Mechanics},
  vol. 641, p. 115–127, 2009.

\bibitem{schmid_2010}
P.~J. SCHMID, ``Dynamic mode decomposition of numerical and experimental
  data,'' \emph{Journal of Fluid Mechanics}, vol. 656, p. 5–28, 2010.

\bibitem{Schmid2011}
\BIBentryALTinterwordspacing
P.~J. Schmid, L.~Li, M.~P. Juniper, and O.~Pust, ``Applications of the dynamic
  mode decomposition,'' \emph{Theoretical and Computational Fluid Dynamics},
  vol.~25, no.~1, pp. 249--259, Jun 2011. [Online]. Available:
  \url{https://doi.org/10.1007/s00162-010-0203-9}
\BIBentrySTDinterwordspacing

\bibitem{BRUNTON20161}
\BIBentryALTinterwordspacing
B.~W. Brunton, L.~A. Johnson, J.~G. Ojemann, and J.~N. Kutz, ``Extracting
  spatial–temporal coherent patterns in large-scale neural recordings using
  dynamic mode decomposition,'' \emph{Journal of Neuroscience Methods}, vol.
  258, pp. 1--15, 2016. [Online]. Available:
  \url{https://www.sciencedirect.com/science/article/pii/S0165027015003829}
\BIBentrySTDinterwordspacing

\bibitem{b32}
\BIBentryALTinterwordspacing
F.~Salfner, M.~Lenk, and M.~Malek, ``A survey of online failure prediction
  methods,'' \emph{ACM Comput. Surv.}, vol.~42, no.~3, Mar. 2010. [Online].
  Available: \url{https://doi.org/10.1145/1670679.1670680}
\BIBentrySTDinterwordspacing

\bibitem{b33}
\BIBentryALTinterwordspacing
D.~Jauk, D.~Yang, and M.~Schulz, ``Predicting faults in high performance
  computing systems: An in-depth survey of the state-of-the-practice,'' in
  \emph{Proceedings of the International Conference for High Performance
  Computing, Networking, Storage and Analysis}, ser. SC '19.\hskip 1em plus
  0.5em minus 0.4em\relax New York, NY, USA: Association for Computing
  Machinery, 2019. [Online]. Available:
  \url{https://doi.org/10.1145/3295500.3356185}
\BIBentrySTDinterwordspacing

\bibitem{surveyreliability}
\BIBentryALTinterwordspacing
S.~He, P.~He, Z.~Chen, T.~Yang, Y.~Su, and M.~R. Lyu, ``A survey on automated
  log analysis for reliability engineering,'' \emph{ACM Comput. Surv.},
  vol.~54, no.~6, jul 2021. [Online]. Available:
  \url{https://doi.org/10.1145/3460345}
\BIBentrySTDinterwordspacing

\bibitem{b13}
Z.~Zheng, Z.~Lan, B.-H. Park, and A.~Geist, ``System log pre-processing to
  improve failure prediction,'' \emph{2009 IEEE/IFIP International Conference
  on Dependable Systems \& Networks}, pp. 572--577, 2009.

\bibitem{b31}
S.~Jain, I.~Singh, A.~Chandra, Z.-L. Zhang, and G.~Bronevetsky, ``Extracting
  the textual and temporal structure of supercomputing logs,'' in \emph{2009
  International Conference on High Performance Computing (HiPC)}, 2009, pp.
  254--263.

\bibitem{9751445}
Shilpika, T.~Fujiwara, N.~Sakamoto, J.~Nonaka, and K.-L. Ma, ``A visual
  analytics approach for hardware system monitoring with streaming functional
  data analysis,'' \emph{IEEE Transactions on Visualization and Computer
  Graphics}, pp. 1--1, 2022.

\bibitem{b8}
\BIBentryALTinterwordspacing
A.~Gainaru, F.~Cappello, and W.~Kramer, ``Taming of the shrew: Modeling the
  normal and faulty behaviour of large-scale hpc systems,'' in
  \emph{Proceedings of the 2012 IEEE 26th International Parallel and
  Distributed Processing Symposium}, ser. IPDPS '12.\hskip 1em plus 0.5em minus
  0.4em\relax USA: IEEE Computer Society, 2012, p. 1168–1179. [Online].
  Available: \url{https://doi.org/10.1109/IPDPS.2012.107}
\BIBentrySTDinterwordspacing

\bibitem{b14}
A.~Gainaru, F.~Cappello, S.~Trausan-Matu, and B.~Kramer, ``Event log mining
  tool for large scale hpc systems,'' in \emph{Proceedings of the 17th
  International Conference on Parallel Processing - Volume Part I}, ser.
  Euro-Par'11.\hskip 1em plus 0.5em minus 0.4em\relax Berlin, Heidelberg:
  Springer-Verlag, 2011, p. 52–64.

\bibitem{9825952}
S.~Shilpika, B.~Lusch, M.~Emani, F.~Simini, V.~Vishwanath, M.~E. Papka, and
  K.-L. Ma, ``Toward an in-depth analysis of multifidelity high performance
  computing systems,'' in \emph{2022 22nd IEEE International Symposium on
  Cluster, Cloud and Internet Computing (CCGrid)}, 2022, pp. 716--725.

\bibitem{FUJIWARA201898}
\BIBentryALTinterwordspacing
T.~Fujiwara, J.~K. Li, M.~Mubarak, C.~Ross, C.~D. Carothers, R.~B. Ross, and
  K.-L. Ma, ``A visual analytics system for optimizing the performance of
  large-scale networks in supercomputing systems,'' \emph{Visual Informatics},
  vol.~2, no.~1, pp. 98--110, 2018, proceedings of PacificVAST 2018. [Online].
  Available:
  \url{https://www.sciencedirect.com/science/article/pii/S2468502X18300160}
\BIBentrySTDinterwordspacing

\bibitem{8048931}
J.~K. Li, M.~Mubarak, R.~B. Ross, C.~D. Carothers, and K.-L. Ma, ``Visual
  analytics techniques for exploring the design space of large-scale high-radix
  networks,'' in \emph{2017 IEEE International Conference on Cluster Computing
  (CLUSTER)}, 2017, pp. 193--203.

\bibitem{shilpika2019mela}
F.~Shilpika, B.~Lusch, M.~Emani, V.~Vishwanath, M.~E.~Papka, and K.-L. Ma,
  ``Mela: A visual analytics tool for studying multifidelity hpc system logs,''
  in \emph{2019 IEEE/ACM Industry/University Joint International Workshop on
  Data-center Automation, Analytics, and Control (DAAC)}, 2019, pp. 13--18.

\bibitem{8585646}
T.~Fujiwara, P.~Malakar, K.~Reda, V.~Vishwanath, M.~E. Papka, and K.-L. Ma, ``A
  visual analytics system for optimizing communications in massively parallel
  applications,'' in \emph{2017 IEEE Conference on Visual Analytics Science and
  Technology (VAST)}, 2017, pp. 59--70.

\bibitem{8973380}
J.~K. Li, T.~Fujiwara, S.~P. Kesavan, C.~Ross, M.~Mubarak, C.~D. Carothers,
  R.~B. Ross, and K.-L. Ma, ``A visual analytics framework for analyzing
  parallel and distributed computing applications,'' in \emph{2019 IEEE
  Visualization in Data Science (VDS)}, 2019, pp. 1--9.

\bibitem{10.1145/3411764.3445396}
\BIBentryALTinterwordspacing
X.~Zhang, S.~Chandrasegaran, and K.-L. Ma, ``Conceptscope: Organizing and
  visualizing knowledge in documents based on domain ontology,'' in
  \emph{Proceedings of the 2021 CHI Conference on Human Factors in Computing
  Systems}, ser. CHI '21.\hskip 1em plus 0.5em minus 0.4em\relax New York, NY,
  USA: Association for Computing Machinery, 2021. [Online]. Available:
  \url{https://doi.org/10.1145/3411764.3445396}
\BIBentrySTDinterwordspacing

\bibitem{255601}
\BIBentryALTinterwordspacing
M.~Brundage, S.~Chandrasegaran, X.~Zhang, and K.-L. Ma,
  ``\BIBforeignlanguage{en}{Using text visualization to aid analysis of machine
  maintenance logs}.''\hskip 1em plus 0.5em minus 0.4em\relax Proceedings of
  the 11th Model-Based Enterprise Summit, Gaithersburg, MD, 2020-04-30 2020.
  [Online]. Available:
  \url{https://tsapps.nist.gov/publication/get_pdf.cfm?pub_id=930322}
\BIBentrySTDinterwordspacing

\bibitem{8901998}
H.~T. Nguyen, A.~Bhatele, N.~Jain, S.~P. Kesavan, H.~Bhatia, T.~Gamblin, K.-L.
  Ma, and P.-T. Bremer, ``Visualizing hierarchical performance profiles of
  parallel codes using callflow,'' \emph{IEEE Transactions on Visualization and
  Computer Graphics}, vol.~27, no.~4, pp. 2455--2468, 2021.

\bibitem{9622132}
S.~Kesavan, H.~Bhatia, A.~Bhatele, S.~Brink, O.~Pearce, T.~Gamblin, P.-T.
  Bremer, and K.-L. Ma, ``Scalable comparative visualization of ensembles of
  call graphs,'' \emph{IEEE Transactions on Visualization and Computer
  Graphics}, pp. 1--1, 2021.

\bibitem{b2}
\BIBentryALTinterwordspacing
S.~Di, R.~Gupta, M.~Snir, E.~Pershey, and F.~Cappello, ``Logaider: A tool for
  mining potential correlations of hpc log events,'' in \emph{Proceedings of
  the 17th IEEE/ACM International Symposium on Cluster, Cloud and Grid
  Computing}, ser. CCGrid '17.\hskip 1em plus 0.5em minus 0.4em\relax IEEE
  Press, 2017, p. 442–451. [Online]. Available:
  \url{https://doi.org/10.1109/CCGRID.2017.18}
\BIBentrySTDinterwordspacing

\bibitem{b15}
\BIBentryALTinterwordspacing
C.~D. Martino, S.~Jha, W.~Kramer, Z.~Kalbarczyk, and R.~K. Iyer, ``Logdiver: A
  tool for measuring resilience of extreme-scale systems and applications,'' in
  \emph{Proceedings of the 5th Workshop on Fault Tolerance for HPC at EXtreme
  Scale}, ser. FTXS '15.\hskip 1em plus 0.5em minus 0.4em\relax New York, NY,
  USA: Association for Computing Machinery, 2015, p. 11–18. [Online].
  Available: \url{https://doi.org/10.1145/2751504.2751511}
\BIBentrySTDinterwordspacing

\bibitem{b16}
\BIBentryALTinterwordspacing
H.~Hamooni, B.~Debnath, J.~Xu, H.~Zhang, G.~Jiang, and A.~Mueen, ``Logmine:
  Fast pattern recognition for log analytics,'' in \emph{Proceedings of the
  25th ACM International on Conference on Information and Knowledge
  Management}, ser. CIKM '16.\hskip 1em plus 0.5em minus 0.4em\relax New York,
  NY, USA: Association for Computing Machinery, 2016, p. 1573–1582. [Online].
  Available: \url{https://doi.org/10.1145/2983323.2983358}
\BIBentrySTDinterwordspacing

\bibitem{b18}
X.~Fu, R.~Ren, J.~Zhan, W.~Zhou, Z.~Jia, and G.~Lu, ``Logmaster: Mining event
  correlations in logs of large-scale cluster systems,'' in \emph{2012 IEEE
  31st Symposium on Reliable Distributed Systems}, 2012, pp. 71--80.

\bibitem{b92}
\BIBentryALTinterwordspacing
A.~Netti, M.~M\"{u}ller, C.~Guillen, M.~Ott, D.~Tafani, G.~Ozer, and M.~Schulz,
  ``Dcdb wintermute: Enabling online and holistic operational data analytics on
  hpc systems,'' in \emph{Proceedings of the 29th International Symposium on
  High-Performance Parallel and Distributed Computing}, ser. HPDC '20.\hskip
  1em plus 0.5em minus 0.4em\relax New York, NY, USA: Association for Computing
  Machinery, 2020, p. 101–112. [Online]. Available:
  \url{https://doi.org/10.1145/3369583.3392674}
\BIBentrySTDinterwordspacing

\bibitem{kesavan2020visual}
S.~P. Kesavan, T.~Fujiwara, J.~K. Li, C.~Ross, M.~Mubarak \emph{et~al.}, ``A
  visual analytics framework for reviewing streaming performance data,'' in
  \emph{Proc. PacificVis}, 2020, pp. 206--215.

\bibitem{b11}
\BIBentryALTinterwordspacing
G.~Lakner and B.~Knudson, ``Ibm system blue gene solution: Blue gene/q system
  administration.'' [Online]. Available:
  \url{https://www.oreilly.com/library/view/ibm-system-blue/0738438006/}
\BIBentrySTDinterwordspacing

\bibitem{tut}
\BIBentryALTinterwordspacing
J.~H. Tu, C.~W. Rowley, D.~M. Luchtenburg, S.~L. Brunton, and J.~N. Kutz, ``On
  dynamic mode decomposition: Theory and applications,'' \emph{Journal of
  Computational Dynamics}, vol.~1, no.~2, pp. 391--421, 2014. [Online].
  Available: \url{/article/id/1dfebc20-876d-4da7-8034-7cd3c7ae1161}
\BIBentrySTDinterwordspacing

\bibitem{Mann2015DynamicMD}
J.~Mann and J.~N. Kutz, ``Dynamic mode decomposition for financial trading
  strategies,'' \emph{Quantitative Finance}, vol.~16, pp. 1643 -- 1655, 2015.

\bibitem{cntrl}
\BIBentryALTinterwordspacing
J.~L. Proctor, S.~L. Brunton, and J.~N. Kutz, ``Dynamic mode decomposition with
  control,'' \emph{SIAM Journal on Applied Dynamical Systems}, vol.~15, no.~1,
  pp. 142--161, 2016. [Online]. Available:
  \url{https://doi.org/10.1137/15M1013857}
\BIBentrySTDinterwordspacing

\bibitem{Hemati2014DynamicMD}
M.~S. Hemati, M.~O. Williams, and C.~W. Rowley, ``Dynamic mode decomposition
  for large and streaming datasets,'' \emph{Physics of Fluids}, vol.~26, p.
  111701, 2014.

\bibitem{Pendergrass2016StreamingGS}
S.~D. Pendergrass, J.~N. Kutz, and S.~L. Brunton, ``Streaming gpu singular
  value and dynamic mode decompositions,'' \emph{ArXiv}, vol. abs/1612.07875,
  2016.

\bibitem{Dawson2014CharacterizingAC}
S.~T.~M. Dawson, M.~S. Hemati, M.~O. Williams, and C.~W. Rowley,
  ``Characterizing and correcting for the effect of sensor noise in the dynamic
  mode decomposition,'' \emph{Experiments in Fluids}, vol.~57, pp. 1--19, 2014.

\bibitem{Manohar2019OPTIMIZEDSF}
\BIBentryALTinterwordspacing
K.~Manohar, E.~Kaiser, S.~L. Brunton, and J.~N. Kutz, ``Optimized sampling for
  multiscale dynamics,'' \emph{Multiscale Modeling \& Simulation}, vol.~17,
  no.~1, pp. 117--136, 2019. [Online]. Available:
  \url{https://doi.org/10.1137/17M1162366}
\BIBentrySTDinterwordspacing

\bibitem{9524582}
N.~M.~M. Kalimullah, A.~Shelke, and A.~Habib, ``Multiresolution dynamic mode
  decomposition (mrdmd) of elastic waves for damage localisation in
  piezoelectric ceramic,'' \emph{IEEE Access}, vol.~9, pp. 120\,512--120\,524,
  2021.

\bibitem{8709782}
M.~Bilal, M.~Rizwan, S.~Saleem, M.~M. Khan, M.~S. Alkatheir, and M.~Alqarni,
  ``Automatic seizure detection using multi-resolution dynamic mode
  decomposition,'' \emph{IEEE Access}, vol.~7, pp. 61\,180--61\,194, 2019.

\bibitem{b6}
{InfluxData}, ``Influxdb 1.7 documentation,''
  \url{https://docs.influxdata.com/InfluxDB/v1.7}, 2021, accessed: 2021-10-19.

\bibitem{esearch}
C.~Gormley and Z.~Tong, \emph{Elasticsearch: The Definitive Guide},
  1st~ed.\hskip 1em plus 0.5em minus 0.4em\relax O'Reilly Media, Inc., 2015.

\bibitem{b46}
M.~Grinberg, \emph{Flask Web Development: Developing Web Applications with
  Python}, 1st~ed.\hskip 1em plus 0.5em minus 0.4em\relax O'Reilly Media, Inc.,
  2014.

\bibitem{b47}
\BIBentryALTinterwordspacing
M.~Bostock, V.~Ogievetsky, and J.~Heer, ``D3 data-driven documents,''
  \emph{IEEE Transactions on Visualization and Computer Graphics}, vol.~17,
  no.~12, p. 2301–2309, dec 2011. [Online]. Available:
  \url{https://doi.org/10.1109/TVCG.2011.185}
\BIBentrySTDinterwordspacing

\bibitem{numlin}
L.~N. Trefethen and D.~Bau, \emph{Numerical Linear Algebra}.\hskip 1em plus
  0.5em minus 0.4em\relax SIAM, Philadelphia, 1997.

\bibitem{opsvht}
M.~Gavish and D.~L. Donoho, ``The optimal hard threshold for singular values is
  $4/\sqrt {3}$,'' \emph{IEEE Transactions on Information Theory}, vol.~60,
  no.~8, pp. 5040--5053, 2014.

\bibitem{nyq}
\BIBentryALTinterwordspacing
L.~Tan and J.~Jiang, ``Chapter 2 - signal sampling and quantization,'' in
  \emph{Digital Signal Processing (Third Edition)}, third edition~ed., L.~Tan
  and J.~Jiang, Eds.\hskip 1em plus 0.5em minus 0.4em\relax Academic Press,
  2019, pp. 13--58. [Online]. Available:
  \url{https://www.sciencedirect.com/science/article/pii/B9780128150719000026}
\BIBentrySTDinterwordspacing

\bibitem{welch1967use}
P.~Welch, ``The use of fast fourier transform for the estimation of power
  spectra: a method based on time averaging over short, modified
  periodograms,'' \emph{IEEE Transactions on audio and electroacoustics},
  vol.~15, no.~2, pp. 70--73, 1967.

\bibitem{10.5555/2386124.2386138}
C.~Mueller, D.~Gregor, and A.~Lumsdaine, ``Distributed force-directed graph
  layout and visualization,'' in \emph{Proceedings of the 6th Eurographics
  Conference on Parallel Graphics and Visualization}, ser. EGPGV '06.\hskip 1em
  plus 0.5em minus 0.4em\relax Goslar, DEU: Eurographics Association, 2006, p.
  83–90.

\bibitem{9355229}
T.~Patel, Z.~Liu, R.~Kettimuthu, P.~Rich, W.~Allcock, and D.~Tiwari, ``Job
  characteristics on large-scale systems: Long-term analysis, quantification,
  and implications,'' in \emph{SC20: International Conference for High
  Performance Computing, Networking, Storage and Analysis}, 2020, pp. 1--17.

\bibitem{Jolliffe2011}
\BIBentryALTinterwordspacing
I.~Jolliffe, \emph{Principal Component Analysis}.\hskip 1em plus 0.5em minus
  0.4em\relax Berlin, Heidelberg: Springer Berlin Heidelberg, 2011, pp.
  1094--1096. [Online]. Available:
  \url{https://doi.org/10.1007/978-3-642-04898-2_455}
\BIBentrySTDinterwordspacing

\bibitem{mcinnes2018umap}
L.~McInnes, J.~Healy, and J.~Melville, ``{UMAP}: Uniform manifold approximation
  and projection for dimension reduction,'' \emph{arXiv:1802.03426}, 2018.

\bibitem{TRUONG2020107299}
\BIBentryALTinterwordspacing
C.~Truong, L.~Oudre, and N.~Vayatis, ``Selective review of offline change point
  detection methods,'' \emph{Signal Processing}, vol. 167, p. 107299, 2020.
  [Online]. Available:
  \url{https://www.sciencedirect.com/science/article/pii/S0165168419303494}
\BIBentrySTDinterwordspacing

\end{thebibliography}
\end{document}